\documentclass[12pt,a4paper]{article}
\usepackage{amsmath,amssymb,bbm}
\usepackage{latexsym}
\usepackage[dvips]{epsfig}
\usepackage{pstricks}
\usepackage{pst-node}
\usepackage[cp850]{inputenc}
\renewcommand{\theequation}{\thesection.\arabic{equation}}
\renewcommand{\thefootnote}{\fnsymbol{footnote}}
\newlength{\extraspace}
\newlength{\extraspaces}
\newcommand{\be}{\begin{equation}}
\newcommand{\ee}{\end{equation}}
 
\newcommand{\ba}{\begin{eqnarray}}
\newcommand{\ea}{\end{eqnarray}}

\newcommand{\bas}{\begin{eqnarray*}}
\newcommand{\eas}{\end{eqnarray*}}
 
\newcounter{subequation}[equation]
\makeatletter

\expandafter\let\expandafter
\reset@font\csname reset@font\endcsname

\def\subeqnarray{\arraycolsep1pt
    \def\@eqnnum\stepcounter##1{\stepcounter{subequation}%
        {\reset@font\rm(\theequation\alph{subequation})}}
\jot5mm     \eqnarray}

\def\subarray{\arraycolsep1pt
    \def\@eqnnum\stepcounter##1{\stepcounter{subequation}%
        {\reset@font\rm(\alph{subequation})}}
\jot5mm     \eqnarray}

\makeatother

\newcommand{\Zom}{\mathbb{Z}}
\newcommand{\one}{\mathbbm{1}}

\newcommand{\R}{\mathbb{R}}

\newcommand{\bra}{\langle}
\newcommand{\ket}{\rangle}
\newcommand{\ra}{\rightarrow}

\newcommand{\is}{ &\! =\! & }
\newcommand{\nonum}{\nonumber \\[1.5mm]}
\newcommand{\sspace}{\makebox[1cm]{ }}
\newcommand{\bspace}{\makebox[2cm]{ }}

\newcommand{\Tr}{{\rm Tr}}

\renewcommand{\th}{{\theta}}
\newcommand{\eps}{\epsilon}
\newcommand{\lb}{\lambda}
\newcommand{\om}{\omega}

\newcommand{\dd}{{\partial}}

\newcommand{\cD}{{\cal D}}

\newcommand{\cJ}{{\cal J}}
\newcommand{\cL}{{\cal L}}

\newcommand{\cM}{{\cal M}}

\newcommand{\cR}{{\cal R}}
\newcommand{\cS}{{\cal S}}

\newcommand{\cZ}{{\cal Z}}

\newcommand{\bg}{\bar{g}}
 \newcommand{\ub}{\bar{u}}

\newcommand{\dk}{ k\frac{\dd}{\dd k} }
\newcommand{\mn}{{\mu\nu}}

\renewcommand{\baselinestretch}{1.2}
\author{Martin Reuter$^1$ and Frank Saueressig$^2$}
\date{} 

\title{Functional Renormalization Group Equations, \\
Asymptotic Safety, and Quantum Einstein Gravity\footnote{Based on lectures given by M.R. at the 
``First Quantum Geometry and Quantum Gravity School'', 
Zakopane, Poland, March 2007, and the 
``Summer School on Geometric and Topological Methods 
for Quantum Field Theory'', Villa de Leyva, Colombia, 
July 2007, and by F.S. at NIKHEF, Amsterdam, The Netherlands, June 2006.}
}

\begin{document}
\maketitle 
\vspace{-8cm}
\begin{flushright} 
MZ-TH/07-05 \\
ITP-UU-07/22 \\
Spin-07/15 \\[4ex]
\end{flushright}
\vspace{6cm}
\centerline{\small\it $^1$ Institute of Physics, University of Mainz, Staudingerweg 7, }
\centerline{\small\it D-55099 Mainz, Germany, E-mail: reuter@thep.physik.uni-mainz.de \\[2ex]} 
\centerline{\small\it $^2$ Institute of Theoretical Physics and Spinoza Institute, Utrecht University,}
\centerline{\small\it  3508 TD Utrecht, Netherlands,
  E-mail: F.S.Saueressig@phys.uu.nl}

\vspace{0.7cm} 

\begin{abstract}
\noindent
These lecture notes provide a pedagogical introduction to a specific continuum implementation of the Wilsonian renormalization group, the effective average action. Its general properties and, in particular, its functional renormalization group equation are explained in a simple scalar setting. The approach is then applied to Quantum Einstein Gravity (QEG). The possibility of constructing a fundamental theory of quantum gravity in the framework of Asymptotic Safety is discussed and the supporting evidence is summarized.
\end{abstract}

\newpage
\renewcommand{\baselinestretch}{1.5}
\Large \normalsize
\renewcommand{\thefootnote}{\arabic{footnote}}
\setcounter{footnote}{0}
%
\section{Introduction}

After the introduction of a functional renormalization 
group equation for gravity \cite{mr} detailed 
investigations of the non-perturbative renormalization 
group (RG) behavior of Quantum Einstein Gravity have 
become possible \cite{mr}-\cite{hier}.
The exact RG equation underlying this approach 
defines a Wilsonian RG flow on a theory space 
which consists of all diffeomorphism invariant 
functionals of the metric $g_{\mu \nu}$. 
The approach turned out to be ideal for 
investigating the asymptotic safety scenario in 
gravity \cite{wein,livrev,QGbooks}
and, in fact, substantial evidence was found for 
the non-perturbative renormalizability
of Quantum Einstein Gravity. The theory emerging 
from this construction (henceforth denoted
``QEG'') is not a quantization of classical 
General Relativity. Instead, its bare action corresponds to
a non-trivial fixed point of the RG flow and is a prediction therefore.
Independent support for the asymptotic safety 
conjecture comes from a two-dimensional 
symmetry reduction of the gravitational path-integral \cite{max}. 

The approach of \cite{mr} employs the effective average
action \cite{avact,ym,avactrev,ymrev} which has crucial advantages as compared to other
continuum implementations of the Wilsonian RG flow \cite{bagber}. 
In particular it is closely related to the standard effective action 
and defines a family of effective field theories $\{ \Gamma_k[g_{\mu \nu}], 0 \le k < \infty \}$ 
labeled by the coarse graining scale $k$. 
The latter property opens the door to a rather direct extraction of physical 
information from the RG flow, at least in single-scale cases: If the physical
 process under consideration involves 
 a single typical momentum scale $p_0$ only, it can be described by a tree-level 
evaluation of $\Gamma_k[g_{\mu \nu}]$, with $k = p_0$.\footnote{ 
The precision which can be achieved by this effective field theory 
description depends on the size of the fluctuations relative 
to mean values. If they turn out large, or if more than one scale is involved, it might be necessary to go beyond the tree-level 
analysis.}

The effective field theory techniques proved useful for an understanding of the scale 
dependent geometry of the effective QEG spacetimes \cite{jan1,jan2,oliverfrac}. In particular it has been shown
\cite{oliver1,oliver2,oliverfrac} that these spacetimes have fractal properties, with
a fractal dimension of 2 at small, and 4 at large distances. The same dynamical dimensional
reduction was also observed in numerical studies of Lorentzian dynamical
triangulations \cite{ajl1,ajl2,ajl34}; in \cite{ncgeom} A.~Connes et al.\ speculated about its possible
relevance to the non-commutative geometry of the standard model.
 
As for possible physics implications of the RG flow 
predicted by QEG, ideas from particle physics, in 
particular the ``RG improvement'', have been employed 
in order to study the leading quantum gravity effects 
in black hole and cosmological spacetimes \cite{bh}-\cite{mof}. 
Among other results, it was found \cite{bh} that the quantum 
effects tend to decrease the Hawking temperature of black holes, 
and that their evaporation process presumably stops 
completely once the black holes mass is of the order of the Planck mass.

These notes are intended to provide the background necessary for understanding these developments.
In the next section we introduce the general idea of the effective average 
action and its associated functional renormalization group equation (FRGE)
 by means of a simple scalar example 
\cite{avact,avactrev}, before reviewing the corresponding 
construction for gravity \cite{mr} in section 3. In all practical calculations 
based upon this approach which have been performed to date
the truncation of theory space has been used as a non-perturbative 
approximation scheme. In section 3 we explain the general ideas 
and problems behind this method, and in section 4 we illustrate it
explicitly in a simple context, the so-called Einstein-Hilbert 
truncation. Section 5 introduces the concept of asymptotic safety while section 6 
contains a summary of the results obtained
using truncated flow equations, with an emphasis on the question
as to whether there exists a non-trivial fixed point for the average action's
RG flow. If so, QEG could be established as a fundamental theory of
quantum gravity which is non-perturbatively renormalizable and ``asymptotically
safe'' from unphysical divergences.

\section{Introducing the effective average action} 
\setcounter{equation}{0}
\label{sect:2}

In this section we introduce the concept of the effective 
average action \cite{avact,avactrev,ym,ymrev} in the simplest context:
 scalar field theory on flat $d$-dimensional Euclidean space $\R^d$.

\subsection{The basic construction for scalar fields} 

We start by considering a single-component real scalar field
$\chi$: $\R^d \ra \R$ whose 
dynamics is governed by the bare action $S[\chi]$. Typically 
the functional $S$ has the structure 
$ 
S[\chi] = \int \! d^dx \, \Big\{ \frac{1}{2} (\dd_{\mu} \chi)^2 + 
\frac{1}{2} m^2 \chi^2 + \mbox{interactions} \Big\}\,,
$
but we shall not need to assume any specific form of $S$ in the 
following. After coupling $\chi(x)$ to a source $J(x)$ we can 
write down an a priori formal path integral representation 
for the generating functional of the connected Green's functions:
$W[J] = \ln \int \! \cD \chi \exp\{-S[\chi] + \int \! d^dx \, \chi(x) J(x)\}$.
By definition, the (conventional) effective action $\Gamma[\phi]$ is the 
Legendre transform of $W[J]$. It depends on the field 
expectation value $\phi \equiv \bra \chi \ket = 
\delta W[J]/\delta J$ and generates all $1$-particle irreducible 
Greens functions of the theory by multiple functional differentiation 
with respect to $\phi(x)$ and setting $\phi = \phi[J =0]$ thereafter.
 In order to make the functional integral well-defined a UV 
cutoff is needed; for example one could replace 
$\R^d$ by a $d$-dimensional lattice $\Zom^d$. The functional integral $\cD \chi$ would then 
read $\prod_{x \in \Zom^d} d \chi(x)$. In the following we implicitly 
assume such a UV regularization but leave the details unspecified 
and use continuum notation for the fields and their Fourier transforms.

The construction of the effective average action \cite{avact} starts
out from a modified form, $W_k[J]$, of the functional $W[J]$ 
which depends on a variable mass scale $k$. This scale is used to 
separate the Fourier modes of $\chi$ into ``short wave length'' 
and ``long wave length'', depending on whether 
or not their momentum square $p^2\equiv p_{\mu} p^{\mu}$ is larger 
or smaller than $k^2$. By construction, the modes with 
$p^2 > k^2$ contribute without any suppression to the 
functional integral defining 
$W_k[J]$, while those with $p^2 < k^2$ contribute only with a reduced 
weight or are suppressed altogether, depending on which variant of the 
formalism is used. The new functional $W_k[J]$ is obtained from the
conventional one by adding a ``cutoff action'' $\Delta_k S[\chi]$ to 
the bare action $S[\chi]$:
\be 
\exp\left\{ W_k[J] \right\} = \int \! \cD \chi \exp\Big\{ - S[\chi] - 
\Delta_k S[\chi] + \int \! d^dx \, \chi(x) J(x) \Big\}\,. 
\label{E2}
\ee 
The factor $\exp\{- \Delta_k S[\chi]\}$ serves the purpose of suppressing 
the ``IR modes'' having $p^2 < k^2$. In momentum space the 
cutoff action is taken to be of the form 
\be 
\Delta_k S[\chi] \equiv \frac{1}{2} \int \! \frac{d^dp}{(2\pi)^d} \,
\cR_k(p^2) \, |\widehat{\chi}(p)|^2\,,
\label{E3}
\ee  
where $\widehat{\chi}(p) = \int \! d^dx\, \chi(x) \exp( - i p x)$ 
is the Fourier transform of $\chi(x)$. The precise shape of the function 
$\cR_k(p^2)$ is arbitrary to some extent; what matters is its 
limiting behavior for $p^2 \gg k^2$ and $p^2 \ll k^2$ only. 
In the simplest case%
\footnote{We shall discuss a slight generalization of these conditions at
the end of this section.}
we require that 
\be
\cR_k(p^2) \approx \left\{ 
\begin{array}{ll} k^2 \quad & \mbox{for} \;\;p^2 \ll k^2\,,\\
 0 \quad & \mbox{for} \;\;p^2 \gg k^2\,.  
\end{array}
\right. 
\label{E4}
\ee 
The first condition leads to a suppression of the 
small momentum modes by a soft mass-like IR cutoff, 
the second guarantees that the large momentum modes are 
integrated out in the usual way. Adding $\Delta_kS$ to the bare 
action $S[\chi]$ leads to 
\be 
S + \Delta_k S = \frac{1}{2} \int\! \frac{d^dp}{(2\pi)^d} 
\Big[p^2 + m^2 + \cR_k(p^2) \Big] \, |\widehat{\chi}(p)|^2 + \mbox{interactions}\,.
\label{E5}
\ee 
Obviously the cutoff function $\cR_k(p^2)$ has the 
interpretation of a momentum dependent mass square which vanishes 
for $p^2 \gg k^2$ and assumes the constant value $k^2$ for 
$p^2 \ll k^2$. How $\cR_k(p^2)$ is assumed to interpolate 
between these two regimes is a matter of calculational 
convenience. In practical calculations one often uses the 
exponential cutoff $\cR_k(p^2) = p^2[\exp(p^2/k^2) -1]^{-1}$, 
but many other choices are possible 
\cite{avactrev,opt}.
One could also think of suppressing the $p^2 < k^2$ modes completely.
This could be achieved by allowing $\cR_k(p^2)$ to diverge for $p^2 \ll k^2$ 
so that $\exp\{-\Delta_k S[\chi]\} \ra 0$ for modes with 
$p^2 \ll k^2$. While this behavior of $\cR_k(p^2)$ 
seems most natural from the viewpoint of a Kadanoff-Wilson type 
coarse graining, its singular behavior makes the resulting generating 
functional  problematic to deal with technically. For this reason, and 
since it still allows for the derivation of an exact RG equation, one 
usually prefers to work with a smooth cutoff satisfying (\ref{E4}). At the 
non-perturbative path integral level it suppresses the long 
wavelength modes by a factor $\exp\{-\frac{1}{2} k^2 \int 
|\widehat{\chi}|^2\}$.
In perturbation theory, according to eq.\ (\ref{E5}), the $\Delta_k S$ term 
leads to the modified propagator $[p^2 + m^2 + \cR_k(p^2)]^{-1}$ 
which equals $[p^2 + m^2 + k^2]^{-1}$ for $p^2 \ll k^2$. 
Thus, when computing loops with this propagator, $k^2$ acts indeed as a 
conventional IR cutoff if $m^2 \ll k^2$. (It plays no role in the 
opposite limit $m^2 \gg k^2$ in which the physical particle 
mass cuts off the $p$-integration.)  We note that by replacing 
$p^2$ with $- \dd^2$ in the argument of $\cR_k(p^2)$ 
the cutoff action can be written in a way which makes no reference 
to the Fourier decomposition of $\chi$:
\be 
\Delta_k S[\chi] = 
\frac{1}{2} \int \! d^dx \, \chi(x) \cR_k(-\dd^2) \chi(x)\,.
\label{E6}
\ee

The next steps towards the definition of the effective average action 
are similar to the usual procedure. One defines the (now $k$-dependent) 
field expectation value $\phi(x) \equiv \bra \chi(x)\ket = 
\delta W_k[J]/\delta J(x)$, assumes that the functional relationship
$\phi  = \phi[J]$ can be inverted to yield $J = J[\phi]$, and 
introduces the Legendre transform of $W_k$, 
\be 
\widetilde{\Gamma}_k[\phi] \equiv \int \! d^dx\,  J(x) \phi(x) 
- W_k[J]\,,
\label{E7}
\ee 
where $J = J[\phi]$. The actual effective average action, denoted by 
$\Gamma_k[\phi]$, is obtained from $\widetilde{\Gamma}_k$ by 
subtracting $\Delta_k S[\phi]$: 
\be 
\Gamma_k[\phi] \equiv \widetilde{\Gamma}_k[\phi] - 
\frac{1}{2} \int \! dx\, \phi(x) \cR_k(-\dd^2) \phi(x)\,.
\label{E8}
\ee
The rationale for this definition becomes clear when we look at the 
list of properties enjoyed by the functional $\Gamma_k$:

\noindent
{\bf (1)} The scale dependence of $\Gamma_k$ is governed by the 
 FRGE 
\be 
\dk \Gamma_k[\phi] = \frac{1}{2} {\rm Tr}\Big[ 
\dk \cR_k \, \Big( \Gamma^{(2)}_k[\phi] + \cR_k\Big)^{-1} \Big]\,.
\label{E9}
\ee
Here the RHS uses a compact matrix notation. 
In a position space 
representation $\Gamma_k^{(2)}$ has the matrix elements 
$\Gamma_k^{(2)}(x,y) \equiv \delta^2 \Gamma_k/\delta \phi(x) \delta \phi(y)$,
i.e., it is the Hessian of the average action, 
$\cR_k(x,y) \equiv \cR_k(-\partial_x^2) \delta(x-y)$, and the 
 trace 
${\rm Tr}$ corresponds to an integral $\int \! d^d x$. 
In \eqref{E9} the implicit UV cutoff can be removed trivially.
This is most easily 
seen in the momentum representation where $\dk \cR_k(p^2)$,
 considered a function of $p^2$, is significantly different 
from zero only in the region centered around $p^2 = k^2$. 
Hence the trace receives contributions from a thin shell 
of momenta $p^2 \approx k^2$ only and is therefore well convergent
both in the UV and IR. 

The RHS of (\ref{E9}) can be rewritten in a style reminiscent
of a one-loop expression:
\be 
\dk \Gamma_k[\phi] = \frac{1}{2} \frac{D}{D \ln k} 
{\rm Tr} \ln \Big( \Gamma^{(2)}_k[\phi] + \cR_k \Big) \,.
\label{E9-1}
\ee
Here the scale derivative $D/D\ln k$ acts only on the $k$-dependence 
of $\cR_k$, not on $\Gamma_k^{(2)}$. The ${\rm Tr}\ln (\cdots)
= \ln \det (\cdots)$ expression in (\ref{E9-1}) differs from 
a standard one-loop determinant in two ways: it 
contains the Hessian of the actual effective action rather than  
that of the bare action $S$ and it has a built in IR 
regulator $\cR_k$. These modifications make (\ref{E9-1}) 
an {\it exact} equation. In a sense, solving it 
amounts to solving the complete theory.

The derivation of (\ref{E9}) proceeds as follows \cite{avact}. 
Taking the $k$-derivative of (\ref{E7}) with (\ref{E2})
and (\ref{E6}) inserted one finds 
\be 
k \frac{\partial}{\partial k} \widetilde{\Gamma}_k[\phi] = \frac{1}{2} 
\int \! d^dx d^dy\, \bra \chi(x) \chi(y)\ket\,\dk \cR_k(x,y) \,,
\label{E10}
\ee
with 
$\bra A \ket \equiv e^{-W_k} \int \! \cD \chi \, A \exp\{- S - \Delta_k S 
- \int J \phi\}$
defining the $J$ and $k$ dependent expectation values. 
Next it is convenient to introduce the connected 
2-point function $G_{xy} \equiv G(x,y) \equiv 
\delta^2  W_k[J]/\delta J(x) \delta J(y)$ and the Hessian 
of $\widetilde{\Gamma}_k$: $(\widetilde{\Gamma}^{(2)}_k)_{xy} 
\equiv \delta^2  \widetilde{\Gamma}_k[J]/\delta \phi(x) \delta \phi(y)$.
Since $W_k$ and $\widetilde{\Gamma}_k$ are related by 
a Legendre transformation one shows in the usual way that 
$G$ and $\widetilde{\Gamma}^{(2)}$ are mutually inverse 
matrices: $G \widetilde{\Gamma}^{(2)} =1$. Furthermore,  taking two $J$-derivatives
of (\ref{E2}) one obtains $\bra \chi(x) \chi(y) \ket 
= G(x,y) + \phi(x) \phi(y)$. Substituting this expression for the 
two-point function into (\ref{E10}) we arrive at 
\be 
\dd_t \widetilde{\Gamma}_k[\phi] = 
\frac{1}{2}{\rm Tr}[\dd_t \cR_k G] + 
\frac{1}{2} \int \! d^d x \, \phi(x) \, \dd_t \cR_k(-\dd^2) \, \phi(x) \,,
\label{E11}
\ee 
where $t \equiv \ln(k/k_0)$. In terms of $\Gamma_k$, the effective average action proper, this 
becomes $\dd_t \Gamma_k[\phi] = \frac{1}{2} {\rm Tr}[\dd_t \cR_k G]$. 
The cancellation of the $\frac{1}{2}\int \phi \cR_k \phi$ term is a 
first motivation for the definition (\ref{E8}) where this 
term is subtracted from the Legendre transform $\widetilde{\Gamma}_k$.
The derivation is completed by noting that $G = [\widetilde{\Gamma}^{(2)}]^{-1} 
= (\Gamma^{(2)}_k + \cR_k)^{-1}$, where the second equality 
follows by differentiating (\ref{E8}): 
$\Gamma_k^{(2)} = \widetilde{\Gamma}_k^{(2)} - \cR_k$. 

\noindent
{\bf (2)} The effective average action satisfies the following 
integro-differential equation: 
\ba 
\exp\{-\Gamma_k[\phi]\} = 
\int \! \cD \chi \, \exp\Big\{ - S[\chi] + \int \! d^dx \, (\chi - \phi) 
\frac{\delta \Gamma_k[\phi]}{\delta \phi} \Big\} \times
\nonum
\times \exp\Big\{ - \int \! 
d^dx \, (\chi - \phi) \cR_k(-\dd^2) (\chi - \phi) \Big\}\,. 
\label{E12}
\ea
This equation is easily derived by combining eqs.\ (\ref{E2}), 
(\ref{E7}) and (\ref{E8}), and by using the effective field equation 
$\delta \widetilde{\Gamma}_k/\delta \phi = J$, which is `dual' 
to $\delta W_k /\delta J = \phi$. (Note that it is 
$\widetilde{\Gamma}_k$ which appears here, not $\Gamma_k$.)

\noindent
{\bf (3)} For $k \ra 0$ the effective average action approaches the ordinary 
effective action, $\lim_{k \rightarrow 0} \Gamma_k = \Gamma$, 
and for $k \ra \infty$ the bare action $\Gamma_{k \ra \infty} = S$. 
The $k \ra 0$ limit is a 
consequence of (\ref{E4}), $\cR_k(p^2)$ vanishes for 
all $ p^2 >0$ when $k \ra 0$. The derivation of the $k \ra \infty$ limit 
makes use of the integro-differential equation (\ref{E12}). A formal 
version the argument is as follows. Since $\cR_k(p^2)$ approaches 
$k^2$ for $k \ra \infty$, the second exponential on the RHS 
of (\ref{E12}) becomes $\exp\{- k^2 \int dx (\chi - \phi)^2\}$, which,
up to a normalization factor, approaches a delta-functional 
$\delta[\chi - \phi]$. The $\chi$ integration can be performed trivially 
then and one ends up with $\lim_{k \ra\infty} \Gamma_k[\phi] = 
S[\phi]$. In a more careful treatment \cite{avact} one shows that the saddle
point approximation of the functional integral in (\ref{E12}) about the 
point $\chi = \phi$ becomes exact in the limit $k \ra \infty$. As a result, 
$\lim_{k \ra \infty} \Gamma_k$ and $S$ differ at most by the infinite mass 
limit of a one-loop determinant, which we suppress here 
since it plays no role in typical applications (see \cite{liouv}        
for a more detailed discussion). 

\noindent
{\bf (4)} The FRGE (\ref{E9}) is independent 
of the bare action $S$ which enters only via the initial condition 
$\Gamma_{\infty} = S$. In the FRGE approach the calculation of the 
path integral for $W_k$ is replaced by integrating the 
RG equation from $k =\infty$, where the initial condition 
$\Gamma_{\infty} = S$ is imposed, down to $k=0$, where the effective 
average action equals the ordinary effective action $\Gamma$, the 
object which we actually would like to know. 

\subsection{Theory space}

\begin{figure}[t]
\leavevmode
\hskip 14mm
\epsfxsize=13cm
\epsfysize=8.9cm
\epsfbox{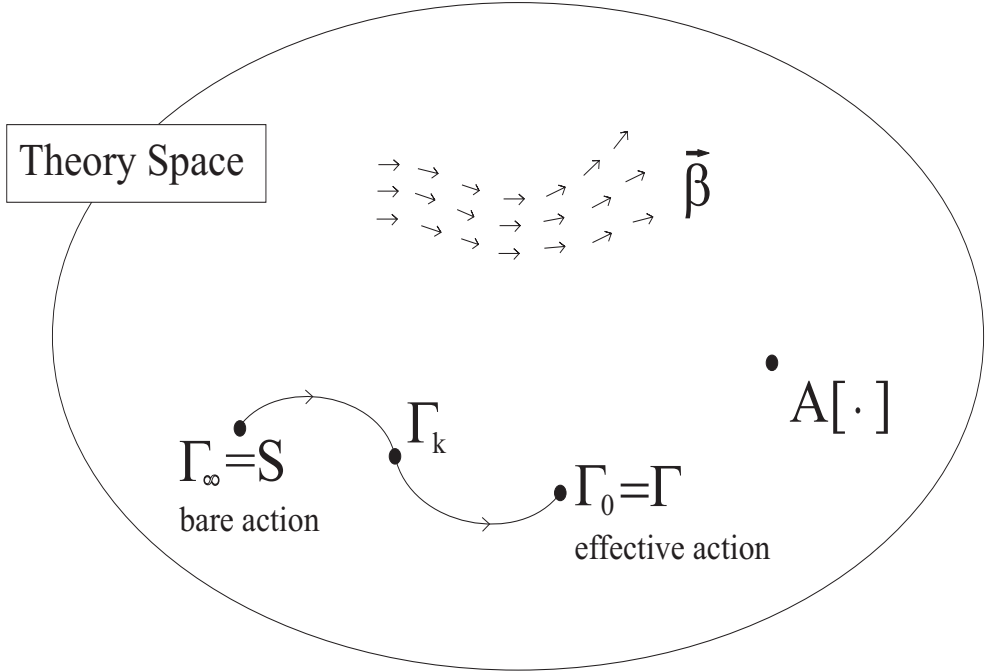}
\vskip 2mm
\caption{\small The points of theory space are the action functionals $A[\, \cdot \,]$. The RG equation defines a vector field $\vec \beta$ on this space; its integral curves are the RG trajectories $k \mapsto \Gamma_k$. They start at the bare action $S$ and end at the standard effective action $\Gamma$.}
\label{theoryspace}
\end{figure}

The arena in which the Wilsonian RG dynamics takes place is the ``theory space''. 
Albeit a somewhat formal notion it helps in visualizing various concepts
related to functional renormalization group equations, see fig.\ \ref{theoryspace}. To describe it,
we shall be slightly more general than in the previous subsection and consider
an arbitrary set of fields $\phi(x)$. Then the corresponding theory space
consists of all (action) functionals $A: \phi \mapsto A[\phi]$ depending on this
set, possibly subject to certain symmetry requirements (a $\Zom_2$-symmetry
for a single scalar, or diffeomorphism  invariance if $\phi$ denotes the spacetime
metric, for instance). So the theory space $\{A[\, \cdot \,]\}$ is fixed once the field 
content and the symmetries are fixed. Let us assume we can find a set of ``basis functionals''
$\{ P_\alpha[ \, \cdot \, ] \}$ so that every point of theory space has an expansion of the form \cite{livrev}
\be\label{Aexpansion}
A[\phi] = \sum_{\alpha = 1}^\infty \, \ub_\alpha \, P_\alpha [\phi]
\ee
The basis $\{ P_\alpha[ \, \cdot \, ] \}$ will include both local field monomials and non-local
invariants and we may use the ``generalized couplings'' $\{ \bar u_\alpha , \alpha = 1,2, \cdots \}$
as local coordinates. More precisely, the theory space is coordinatized by the subset of 
``essential couplings'', i.e., those coordinates which cannot be absorbed by a field reparameterization.

Geometrically speaking the FRGE for the effective average action, eq.\ \eqref{E9} or its generalization for an arbitrary set of fields, defines a vector field $\vec \beta$ on theory space. The integral curves along this vector field are the ``RG trajectories'' $k \mapsto \Gamma_k$ parameterized by the scale $k$. They start, for $k \ra \infty$, at the bare action $S$ (up to the correction term mentioned earlier) and terminate at the ordinary effective action at $k=0$. The natural orientation of the trajectories is from higher to lower scales $k$, the direction of increasing ``coarse graining''. Expanding $\Gamma_k$ as in \eqref{Aexpansion},
\be\label{Gexpansion}
\Gamma_k[\phi] = \sum_{\alpha = 1}^\infty \, \ub_\alpha(k) \, P_\alpha [\phi] \, ,
\ee
the trajectory is described by infinitely many ``running couplings'' $\ub_\alpha(k)$. Inserting \eqref{Gexpansion} into the FRGE we obtain a system of infinitely many coupled differential equations for the $\ub_\alpha$'s:
\be\label{rgeqn1}
k \partial_k \, \ub_\alpha(k) = \overline{\beta}_\alpha(\ub_1 , \ub_2 , \cdots ; k) \; , \quad \alpha = 1,2,\cdots \, .
\ee
Here the ``beta functions'' $\overline{\beta}_\alpha$ arise by 
expanding the trace on the RHS of the FRGE in terms of $\{ P_\alpha[\, \cdot \, ] \}$, i.e.,
$\tfrac{1}{2} \Tr \left[ \cdots \right] = \sum_{\alpha = 1}^\infty \overline{\beta}_\alpha(\ub_1 , \ub_2 , \cdots ; k) P_\alpha[\phi]$. The expansion coefficients $\overline{\beta}_\alpha$ have the 
interpretation of beta functions similar to those of perturbation 
theory, but not restricted to relevant couplings. In standard field theory
jargon one would refer to $\ub_\alpha(k =\infty)$ as the ``bare'' parameters and to 
$\ub_\alpha(k =0)$ as the ``renormalized'' or ``dressed'' parameters. 
\renewcommand{\baselinestretch}{1}
\small\normalsize
\renewcommand{\baselinestretch}{1.5}
\small\normalsize

The notation with the bar on $\ub_\alpha$ and $\overline{\beta}_\alpha$ 
is to indicate that we are still dealing with dimensionful 
couplings. Usually the flow equation is reexpressed in terms of the 
dimensionless couplings $u_\alpha \equiv k^{-d_\alpha} \ub_\alpha$, 
where $d_\alpha$ is the canonical mass dimension of $\ub_\alpha$. Correspondingly the essential $u_\alpha$'s are used as 
coordinates of theory space. The resulting 
RG equations 
\be 
k \dd_k u_\alpha(k) = \beta_\alpha(u_1, u_2, \cdots ) 
\label{E17}
\ee
are a coupled system of autonomous differential equations. 
The $\beta_\alpha$'s have no explicit $k$-dependence and define 
a ``time independent'' vector field on theory space.

Fig.\ \ref{theoryspace} gives a schematic summary of the theory space and its structures.
It should be kept in mind, though, that only the essential couplings are coordinates on theory space, and that $\Gamma_\infty$ and $S$ might differ by a simple, explicitly known functional. 

\subsection{Non-perturbative approximations through truncations}

Up to this point our discussion did not involve any approximation.
In practice, however, it is usually impossible to find exact solutions 
to the flow equation. As a way out, one could evaluate
the trace on the RHS of the FRGE by expanding it with respect to 
some small coupling constant, for instance, thus recovering the familiar 
perturbative beta functions. A more interesting option which gives rise 
to non-perturbative approximate solutions is to truncate the theory 
space $\{A[\,\cdot\,]\}$. The basic idea is to project the RG flow onto a
finite dimensional subspace of theory space. The subspace should be chosen 
in such a way  that the projected flow encapsulates the essential physical 
features of the exact flow on the full space.

Concretely the projection onto a truncation subspace is performed 
as follows. One makes an ansatz of the form 
$ 
\Gamma_k[\phi] = \sum_{i=1}^N {\ub}_i(k) P_i[\phi]\,,
$
where the $k$-independent functionals
$\{P_i[\, \cdot \,], i=1,\cdots,N \}$ form a `basis' on the subspace selected. 
For a scalar field, say, examples include pure potential terms 
$\int d^dx \phi^m(x)$, 
$\int d^dx \phi^n(x) \ln \phi^2(x)$, $\cdots$, a standard kinetic 
term $\int \! d^dx (\dd \phi)^2$, higher order derivative terms 
$\int \! d^dx \, \phi \left({\dd^2} \right)^n \phi$, $\int \! d^d x\, f(\phi) 
\left({\dd^2} \right)^n \phi \,\left({\dd^2} \right)^m \phi$, $\cdots$, and non-local terms like 
$\int \!d^dx \, \phi \ln(-\dd^2) \phi$, $\cdots$.   
Even if $S = \Gamma_{\infty}$ is simple, a standard $\phi^4$ action,
say, the evolution from $k =\infty$ downwards will generate such     
terms, a priori only constrained by symmetry requirements.  
The difficult task in practical RG applications consists in 
selecting a set of $P_i$'s which, on the one hand, is generic enough 
to allow for a sufficiently precise description of the physics one 
is interested in, and which, on the other hand, is small enough to be 
computationally manageable. 

The projected RG flow is described by a set of ordinary (if $N < \infty$) 
differential equations for the couplings $\ub_i(k)$. They arise as follows.
 Let us assume we 
expand the $\phi$-dependence of $\frac{1}{2}{\rm Tr}[\cdots]$ 
(with the ansatz for $\Gamma_k[\phi]$ inserted) in a basis
$\{P_{\alpha}[\, \cdot \,]\}$ of the {\it full} theory space which contains  
the $P_i$'s spanning the truncated space as a subset: 
\be 
\frac{1}{2} {\rm Tr}[\cdots] =  
\sum_{\alpha =1}^{\infty} \overline{\beta}_{\alpha}(\ub_1, \cdots, \ub_N;k) 
\, P_{\alpha}[\phi]     
=      
\sum_{i =1}^N \overline{\beta}_i(\ub_1, \cdots, \ub_N;k) 
\, P_i[\phi] + {\rm rest}\,. 
\label{E15}
\ee
Here the ``rest'' contains all terms outside the truncated theory 
space; the approximation consists in neglecting precisely
those terms. Thus, equating (\ref{E15}) to the LHS of the flow equation,      
$\dd_t \Gamma_k = \sum_{i=1}^N \dd_t \ub_i(k) P_i$, the linear independence 
of the $P_i$'s implies the coupled system of ordinary differential 
equations 
\be 
\dd_t \ub_i(k) = \overline{\beta}_i(\ub_1,\cdots , \ub_N;k)\,,
\quad i = 1, \cdots, N\,.
\label{E16}
\end{equation}
Solving (\ref{E16}) one obtains an {\it approximation} to the 
exact RG trajectory projected onto the chosen subspace. Note that 
this approximate trajectory does, in general, not coincide with 
the projection of the exact trajectory, but if the subspace 
is well chosen, it will not be very different from it. In fact, the most 
non-trivial problem in using truncated flow equations is to
 find and justify a truncation subspace which should be as low dimensional 
as possible to make the calculations feasible, 
but at the same time large enough to describe at least qualitatively 
the essential physics. We shall return to the issue of 
testing the quality of a given truncation later on.

As a simple example of a truncation 
we mention the `local potential approximation' \cite{avactrev}. 
The corresponding subspace consists of functionals containing 
a standard kinetic term plus arbitrary non-derivative terms:
\be 
\Gamma_k[\phi] \equiv \int\! d^d x\, \Big\{ \frac{1}{2} (\dd \phi(x))^2 
+ U_k(\phi(x))\Big\}\,.
\label{E18}
\ee
In this case $N$ is infinite, the coordinates $\ub_i$ on 
truncated theory space being the infinitely many parameters 
characterizing an arbitrary potential function $\phi \mapsto 
U(\phi)$. The infinitely many component equations (\ref{E16}) 
amount to a partial differential equation for the running 
potential $U_k(\phi)$. It is obtained by inserting 
(\ref{E18}) into the FRGE and projecting the trace onto functionals 
of the form (\ref{E18}). This is most easily done by inserting 
a {\it constant} field $\phi =\varphi = const$ into both sides 
of the equation since this gives a non-vanishing value precisely to the 
non-derivative $P_i$'s. Since $\Gamma_k^{(2)} = - \dd^2 + U_k''(\varphi)$ with
$U'' \equiv d^2 U_k/d\phi^2$    
has no explicit $x$-dependence the trace is easily evaluated in 
momentum space. This leads to the following partial differential
equation:   
\be 
k \dd_k U_k(\varphi) = \frac{1}{2}\int \! \frac{d^dp}{(2\pi)^d} 
\frac{k \dd_k \cR_k(p^2)}{ p^2 + \cR_k(p^2) + U_k''(\varphi)}\,.
\label{E19}
\ee
It describes how the classical (or microscopic) potential 
$U_{\infty} = V_{\rm class}$ evolves into the standard effective 
potential $U_0 = V_{\rm eff}$. Remarkably, the limit $\lim_{k \ra 0} U_k$ 
is automatically a convex function of $\varphi$, and there is no need to 
perform the Maxwell construction `by hand', in the case of spontaneous 
symmetry breaking. For a detailed discussion of this 
point we refer to  \cite{avactrev}. 

One can continue the truncation process and make a specific 
ansatz for the $\varphi$-dependence of the running potential, 
$U_k(\varphi) = \frac{1}{2} \overline{m}(k)^2 \varphi^2 + \frac{1}{12} 
\overline{\lb}(k) \varphi^4$, say. Then, upon inserting 
$U''_k(\varphi) = \overline{m}(k)^2 + \overline{\lb}(k) \varphi^2$ into the 
RHS of (\ref{E18}) and expanding to $O(\varphi^4)$ one can 
equate the coefficients of $\varphi^2$ and $\varphi^4$ to obtain 
the flow equations on a 2-dimensional subspace: 
$k \dd_k \overline{m}^2= \overline{\beta}_{\overline{m}^2}$, $k \dd_k \overline{\lb} = 
\overline{\beta}_{\overline{\lb}}$. 

If one wants to go beyond the local potential approximation 
(\ref{E18}) the first step is to allow for a ($\phi$ independent
in the simplest case) wave function renormalization, i.e., a running 
prefactor of the kinetic term: $\Gamma_k = 
\int \! d^dx \, \{ \frac{1}{2} Z_k (\dd \phi)^2 + U_k\}$. 
Using truncations of this type one should employ a slightly different 
normalization of $\cR_k(p^2)$, namely $\cR_k(p^2) \approx 
Z_k k^2$ for $p^2 \ll k^2$. Then $\cR_k$ combines with $\Gamma_k^{(2)}$ 
to the inverse propagator $\Gamma_k^{(2)} + \cR_k = 
Z_k( p^2 + k^2) + \cdots$, as it is necessary if the IR cutoff 
is to give rise to a $({\rm mass})^2$ of size $k^2$ rather than 
$k^2/Z_k$. In particular in more complicated theories 
with more than one field it is important that all fields are 
cut off at precisely the same $k^2$. This is achieved by 
a cutoff function of the form 
\be 
\cR_k(p^2) = \cZ_k \, k^2 \, R^{(0)}(p^2/k^2)\,, 
\label{E20}
\ee 
where $R^{(0)}$ is normalized such that $R^{(0)}(0)=1$ 
and $R^{(0)}(\infty) =0$. In general the factor $\cZ_k$ is 
a matrix in field space. In the sector of modes with inverse 
propagator $Z_k^{(i)} p^2 + \cdots$ this matrix is chosen 
diagonal with entries $\cZ_k = Z_k^{(i)}$.

\section{The effective average action for gravity}
\setcounter{equation}{0}
We saw that the FRGE of the effective average action does not 
depend on the bare action $S$. Given a theory space, the form of the 
FRGE and, as a result, the vector field $\vec \beta$ are completely fixed. 
 To define a theory space $\{A[\,\cdot\,]\}$ 
one has to specify on which types of fields the functionals 
$A$ are supposed to depend, and what their symmetries are.
This is the only input data needed for finding the RG flow. 

In the case of QEG the theory space consists, by 
definition, of functionals
$A[g_{\mu\nu}]$ depending on a symmetric tensor 
field, the metric, in a diffeomorphism invariant way.   
Unfortunately it is not possible to straightforwardly apply the 
constructions of the previous section to this theory space. 
Diffeomorphism invariance leads to two types of complications one 
has to deal with \cite{mr}. 

The first one is not specific to the RG approach. 
It occurs already in the standard functional integral 
quantization of gauge or gravity theories, and is familiar 
from Yang-Mills theories. If one gauge-fixes the functional 
integral with an ordinary (covariant) gauge fixing condition
like $\dd^{\mu} A_{\mu}^a =0$, couples the (non-abelian) gauge field $A_{\mu}^a$ to a source, 
and constructs the ordinary effective action the resulting 
functional $\Gamma[A_{\mu}^a]$ is {\it not} invariant under 
the gauge transformations of $A_{\mu}^a$, $A_{\mu}^a \mapsto 
A_{\mu}^a + D_{\mu}^{ab}(A) \, \om^b$. Only at the level of physical   
quantities constructed from $\Gamma[A_{\mu}^a]$, 
S-matrix elements for instance, gauge invariance is recovered. 

The second problem is related to the fact that in a gauge 
theory a ``coarse graining'' based on a naive Fourier decomposition 
of $A_{\mu}^a(x)$ is not gauge covariant and hence not physical.
In fact, if one were to gauge transform a slowly varying 
$A_{\mu}^a(x)$ using a parameter function $\om^a(x)$ with a fast 
$x$-variation, a gauge field with a fast $x$-variation would arise
which, however, still describes the same physics. 
In a non-gauge theory the coarse graining is performed by 
expanding the field in terms of eigenfunctions of the (positive) 
operator $-\dd^2$ and declaring its eigenmodes `long' or `short' 
wavelength depending on whether the corresponding eigenvalue $p^2$ is smaller 
or larger than a given $k^2$. In a gauge theory the best one can do 
in installing this procedure is to expand with respect to 
the {\it covariant} Laplacian or a similar operator, and then 
organize the modes according to the size of their eigenvalues.
While gauge covariant, this approach sacrifices to some extent 
the intuition of a Fourier coarse graining in terms of slow and fast modes. 
Analogous remarks apply to theories of gravity covariant under 
general coordinate transformations. 

The key idea which led to a solution of both problems
was the use of the background field method. In fact, it is 
well kown \cite{back,joos} that the background gauge fixing 
method leads to an effective action which depends on its arguments in a gauge
invariant way. As it turned out \cite{ym,mr} this technique also lends 
itself for implementing a covariant IR cutoff, and it is at the core 
of the effective average action for Yang-Mills theories \cite{ym,ymrev} and 
for gravity \cite{mr}. In the following we briefly  review the 
effective average action for gravity which has been introduced in ref.\ \cite{mr}. 

The ultimate goal is to give meaning to an integral over `all' 
metrics $\gamma_{\mu\nu}$ of the form $\int \! \cD \gamma_{\mu\nu} \,
\exp\{ - S[\gamma_{\mu\nu}] + {\rm source \; terms}\}$ whose 
bare action $S[\gamma_{\mu\nu}]$ is invariant under general 
coordinate transformations,
\be 
\delta \gamma_{\mu\nu} = \cL_v \gamma_{\mu \nu} \equiv
v^{\rho} \dd_{\rho} \gamma_{\mu\nu} 
+ \dd_{\mu} v^{\rho} \gamma_{\rho \nu} + 
\dd_{\nu} v^{\rho} \gamma_{\rho \mu} \,,
\label{F1}
\ee
where $\cL_v$ is the Lie derivative with respect to the vector 
field $v^{\mu}\dd_{\mu}$. To start with we consider $\gamma_{\mu\nu}$
to be a Riemannian metric and assume that $S[\gamma_{\mu\nu}]$ is positive 
definite. Heading towards the background field formalism, the 
first step consists in decomposing the variable of integration 
according to $\gamma_{\mu\nu} = \bg_{\mu\nu} + h_{\mu\nu}$, 
where $\bg_{\mu\nu}$ is a fixed background metric. Note that 
we are not implying a perturbative expansion here, $h_{\mu\nu}$ 
is not supposed to be small in any sense. After the background split 
the measure $\cD \gamma_{\mu\nu}$ becomes $\cD h_{\mu\nu}$ 
and the gauge transformations which we have to gauge-fix read 
\be 
\delta h_{\mu\nu} = \cL_v \gamma_{\mu\nu} = 
\cL_v( \bg_{\mu\nu} + h_{\mu\nu})\,,\quad 
\delta \bg_{\mu\nu}= 0\,.
\label{F2}
\ee
Picking an a priori arbitrary gauge fixing condition $F_{\mu}(h;\bg) =0$ 
the Faddeev-Popov trick can be applied straightforwardly \cite{back}. 
Upon including an IR cutoff as in the scalar case we are lead to 
the following $k$-dependent generating functional $W_k$ for the 
connected Green functions:
\ba 
&& \exp\left\{ W_k[t^{\mu\nu},\sigma^{\mu}, \bar{\sigma}_{\mu}; 
\bg_{\mu\nu}] \right\} = \int \! \cD h_{\mu\nu} \cD C^\mu \cD \bar{C}_\mu
\,\exp\Big\{ -S[\bar{g}+h]-S_{\rm gf}[h;\bar{g}]
\nonum
&& \bspace \bspace 
-S_{\rm gh}[h,C,\bar{C};\bar{g}]-\Delta_k S[h,C,\bar{C};\bar{g}]
     -S_{\rm source} \Big\}\,.
\label{F3}
\ea 
Here $S_{\rm gf}$ denotes the gauge fixing term    
\be 
S_{\rm gf}[h;\bg]=\frac{1}{2\alpha}\int \! d^dx 
\sqrt{\bg}\,\bg^{\mn} F_\mu F_\nu\,,
\label{F4}
\ee
and $S_{\rm gh}$ is the action for the corresponding Faddeev--Popov
ghosts $C^\mu$ and $\bar{C}_\mu$:
\be
S_{\rm gh}[h,C,\bar{C};\bg]=
-\kappa^{-1}\int \! d^dx \,\bar{C}_\mu\, \bg^{\mu\nu}
\,
\frac{\partial F_\nu}{\partial h_{\alpha\beta}}
\,\cL_C\left(\bg_{\alpha\beta}+h_{\alpha\beta}\right)\,.
\label{F5}
\ee
The Faddeev--Popov action $S_{\rm gh}$ is obtained along the
same lines as in Yang--Mills theory: one applies a gauge
transformation (\ref{F2}) to $F_{\mu}$ and replaces the 
parameters $v^{\mu}$ by the ghost field $C^{\mu}$. The 
integral over $C^{\mu}$ and $\bar{C}_{\mu}$ exponentiates the 
Faddeev-Popov determinant $\det[\delta F_{\mu}/\delta v^{\nu}]$.
In (\ref{F3}) we coupled $h_{\mu\nu}, \,C^{\mu}$ and $\bar{C}_{\mu}$ 
to sources $t^{\mu\nu},\,\bar{\sigma}_{\mu}$ and $\sigma^{\mu}$,
respectively: 
$
S_{\rm source} =  -\int \!d^dx \, \sqrt{\bg}
\Big\{ t^{\mu\nu} h_{\mu\nu} +\bar{\sigma}_\mu C^\mu +\sigma^\mu 
\bar{C}_\mu \Big\} \,.
$
The $k$ and source dependent expectation values of 
$h_{\mu\nu},\, C^{\mu}$ and $\bar{C}_{\mu}$ are then given by 
\be
\label{F7}
\bar{h}_{\mu\nu} = \frac{1}{\sqrt{\bg}}\frac{\delta W_k}{\delta t^{\mu\nu}}
\qquad , \qquad
\xi^\mu=\frac{1}{\sqrt{\bg}}\frac{\delta W_k}{\delta \bar{\sigma}_\mu}
\qquad , \qquad
\bar{\xi}_\mu=\frac{1}{\sqrt{\bg}}\frac{\delta W_k}{\delta\sigma^\mu}\,.
\ee
As usual we assume that one can invert the relations (\ref{F7}) 
and solve for the sources $(t^{\mu\nu}\,,\, \sigma^\mu \, , \, 
\bar{\sigma}_\mu )$ as functionals of 
$(\bar{h}_{\mu\nu} \, , \, \xi^\mu \, , \, \bar{\xi}_\mu )$ and,
parameterically, of $\bg_{\mu\nu}$. The Legendre transform 
$\widetilde{\Gamma}_k$ of $W_k$ reads 
\be
\label{F8}
\widetilde{\Gamma}_k[\bar{h},\xi,\bar{\xi}; \bg]
= \int \! d^dx \, \sqrt{\bg}
\left\{ t^{\mu\nu} \bar{h}_{\mu\nu} +
\bar{\sigma}_\mu \xi^\mu + \sigma^\mu\bar{\xi}_\mu
\right\} -W_k[t,\sigma,\bar{\sigma}; \bg]\,.
\ee
This functional inherits a parametric $\bg_{\mu\nu}$-dependence from 
$W_k$. 

As mentioned earlier for a generic gauge fixing condition the 
Legendre transform (\ref{F8}) is not a diffeomorphism invariant  
functional of its arguments since the gauge breaking under the 
functional integral is communicated to $\widetilde{\Gamma}_k$ via
the sources. While  $\widetilde{\Gamma}_k$ does indeed describe the correct
`on-shell' physics satisfying all constraints coming from BRST invariance, it is not invariant off-shell 
\cite{back,joos}. The situation is different for the class of 
gauge fixing conditions of the background type. While --  
as any gauge fixing condition must -- they break the invariance under 
(\ref{F2}) they are  chosen to be invariant under the so-called
background gauge transformations 
\be 
\delta h_{\mu\nu} = \cL_v h_{\mu\nu} \,,\sspace 
\delta \bg_{\mu\nu} = \cL_v \bg_{\mu\nu} \,. 
\label{F9}
\ee
The complete metric $\gamma_{\mu\nu} = g_{\mu\nu} + h_{\mu\nu}$ 
transforms as $\delta \gamma_{\mu\nu} = \cL_v \gamma_{\mu\nu}$ both 
under (\ref{F9}) and under (\ref{F2}). The crucial difference 
is that the (`quantum') gauge transformations (\ref{F2}) keep $\bg_{\mu\nu}$
 unchanged  so that the entire 
change of $\gamma_{\mu\nu}$ is ascribed to $h_{\mu\nu}$. This is 
the point of view one adopts in a standard perturbative calculation 
around flat space where one fixes $\bg_{\mu\nu} = \eta_{\mu\nu}$ and 
allows for no variation of the background. In the present 
construction, instead, we leave $\bg_{\mu\nu}$ unspecified but insist on
covariance under (\ref{F9}). This will lead to a completely 
background covariant formulation.        

Clearly there exist many possible gauge fixing terms 
$S_{\rm gf}[h;\bg]$ of the form (\ref{F4}) which break 
(\ref{F2}) and are invariant under (\ref{F9}). A convenient 
choice which has been employed in practical calculations 
is the background version of the harmonic coordinate condition 
\cite{back}: 
\be 
F_{\mu} = \sqrt{2} \kappa \Big[\delta_{\mu}^{\beta} 
\bg^{\alpha \gamma} \bar{D}_{\gamma} - 
\frac{1}{2} \bg^{\alpha \beta} \bar{D}_{\mu} \Big] \, h_{\alpha\beta} \,.
\label{F10}
\ee
The covariant derivative $\bar{D}_{\mu}$ involves the Christoffel 
symbols $\bar{\Gamma}^{\rho}_{\mu\nu}$ of the background metric. 
Note that (\ref{F10}) is linear in the quantum field 
$h_{\alpha \beta}$. On a flat background with $\bg_{\mu\nu} = \eta_{\mu\nu}$ 
the condition $F_{\mu} =0$ reduces to the familiar harmonic coordinate 
condition, $\dd^{\mu} h_{\mu\nu} = 
\frac{1}{2} \dd_{\nu} h_{\mu}^{\; \mu}$. In eqs.\ (\ref{F10}) and (\ref{F5}) 
$\kappa$ is an arbitrary constant with the dimension of a mass. We shall 
set $\kappa \equiv (32 \pi \bar{G})^{-1/2}$ with $\bar{G}$ a constant reference value of 
Newton's constant. The ghost action for the gauge 
condition (\ref{F10}) reads 
\be
\label{F11a}
S_{\rm gh}[h,C,\bar{C};\bg]=-\sqrt{2}\int \!d^dx \,  \sqrt{\bg}
\,\bar{C}_\mu \cM [g,\bg]^\mu{}_\nu C^\nu
\ee
with the Faddeev--Popov operator
\be
\label{F11b}
\cM[g,\bg]^\mu{}_\nu=
\bg^{\mu\rho} \bg^{\sigma\lambda}
\bar{D}_\lambda(g_{\rho\nu} D_\sigma +g_{\sigma\nu}D_\rho)
-\bg^{\rho\sigma}\bg^{\mu\lambda}\bar{D}_\lambda g_{\sigma\nu} 
D_\rho\,.
\ee   
It will prove crucial that for every background-type 
choice of $F_{\mu}$, $S_{\rm gh}$ is invariant under (\ref{F9}) together with 
\be 
\delta C^{\mu} = \cL_v C^{\mu} \,,\sspace 
\delta \bar{C}_{\mu} = \cL_v \bar{C}_{\mu} \,. 
\label{F11c}
\ee

The essential piece in eq.~(\ref{F3}) is the IR cutoff for the gravitational
field $h_{\mu\nu}$ and for the ghosts. It is taken to be of the form
\be
\label{F12}
\Delta_k S
= \frac{\kappa^2}{2}\int\! d^dx \, \sqrt{\bg}\, h_{\mu\nu}
\cR^{\rm grav}_k[\bg]^{\mu\nu \rho\sigma}h_{\rho\sigma}
  +\sqrt{2}\int d^dx\, \sqrt{\bg}\, \bar{C}_\mu \cR^{\rm gh}_k[\bg]C^\mu\,.
\ee
The cutoff operators $\cR^{\rm grav}_k$ and $\cR^{\rm gh}_k$ serve the purpose
of discriminating between high--momentum and low--momentum modes.
Eigenmodes of $-\bar{D}^2$ with eigenvalues $p^2\gg k^2$ are integrated out
 without any suppression whereas modes with small eigenvalues
$p^2\ll k^2$ are suppressed.
The operators $\cR^{\rm grav}_k$ and $\cR^{\rm gh}_k$ have the structure
$
\cR_k[\bg]=\cZ_k k^2 R^{(0)}(-\bar{D}^2/k^2)\,,
$
where the dimensionless function $R^{(0)}$ interpolates between $R^{(0)}(0)=1$ and $R^{(0)}(\infty)=0$.
A convenient choice is, e.g., the exponential cutoff 
$R^{(0)}(w)=w[\exp(w)-1]^{-1}$ where $w = p^2/k^2$.
The factors $\cZ_k$ are different for the graviton and the ghost cutoff.
For the ghost $\cZ_k \equiv Z^{\rm gh}_k$ is a pure number, whereas for
the metric fluctuation $\cZ_k \equiv \cZ^{\rm grav}_k$ is a
tensor, constructed only from the background metric $\bg_{\mu\nu}$,
which must be fixed along the lines described at the end of section \ref{sect:2}. 

A feature of $\Delta_k S$ which is essential from a practical point of view 
is that the modes of $h_{\mu\nu}$ and the ghosts are organized according 
to their eigenvalues with respect to the {\it background} Laplace 
operator $\bar{D}^2 = \bg^{\mu\nu} \bar{D}_{\mu} \bar{D}_{\nu}$ rather 
than $D^2 = g^{\mu\nu} D_{\mu} D_{\nu}$, which would pertain to the full 
quantum metric $\bg_{\mu\nu} + h_{\mu\nu}$. Using $\bar{D}^2$ the 
functional $\Delta_k S$ is quadratic in the quantum field $h_{\mu\nu}$,
while it becomes extremely complicated if $D^2$ is used instead. 
The virtue of a quadratic $\Delta_k S$ is that it gives 
rise to a flow equation which contains only {\it second} functional 
derivatives of $\Gamma_k$ but no higher ones. The flow equations resulting 
from the cutoff operator $D^2$ are prohibitively complicated and 
can hardly be used for practical computations. A second property 
of $\Delta_k S$ which is crucial for our purposes is that it is invariant 
under the background gauge transformations (\ref{F9}) with (\ref{F12}). 

Having specified all the ingredients which enter the functional  
integral (\ref{F3}) for the generating functional $W_k$ we can 
write down the final definition of the effective average action 
$\Gamma_k$. It is obtained from the Legendre transform 
$\widetilde{\Gamma}_k$ by subtracting the cutoff action 
$\Delta_k S$ with the classical fields inserted:
\be
\label{F14}
\Gamma_k[\bar{h},\xi,\bar{\xi}; \bg]
= \widetilde{\Gamma}_k[\bar{h},\xi,\bar{\xi}; \bg]
- \Delta_k S[\bar{h}, \xi , \bar{\xi} ; \bg]\,.
\ee
It is convenient to define the expectation value of the quantum metric $\gamma_{\mu \nu}$,
\be
\label{F15}
g_{\mu\nu}(x) \equiv \bg_{\mu\nu}(x) +\bar{h}_{\mu\nu}(x)\,,
\ee
and consider $\Gamma_k$ as a functional of $g_{\mu\nu}$ rather than 
$\bar{h}_{\mu\nu}$:
\be
\label{F16}
\Gamma_k[g_{\mu\nu} ,\bg_{\mu\nu} , \xi^\mu , \bar{\xi}_\mu]
\equiv \Gamma_k[g_{\mu\nu} -\bg_{\mu\nu} , \xi^\mu , \bar{\xi}_\mu; 
\bg_{\mu\nu}]\,.
\ee

So, what did we gain going through this seemingly complicated 
background field construction, eventually ending up 
with an action functional which depends on {\it two} metrics even?
The main advantage of this setting is that the corresponding 
functionals $\widetilde{\Gamma}_k$, and as a result $\Gamma_k$, 
are  invariant under general coordinate transformations
where all its arguments transform as tensors of the corresponding rank:
\be
\label{F17}
\Gamma_k[\Phi +\cL_v \Phi] = \Gamma_k[\Phi]\,,
\qquad  \qquad
\Phi \equiv \left\{
  g_\mn , \bg_\mn , \xi^\mu , \bar{\xi}_\mu \right\}\,.
\ee
Note that in (\ref{F17}), contrary to the ``quantum gauge transformation''
(\ref{F2}), also the background metric transforms as an ordinary
tensor field: $\delta\bg_\mn = \cL_v \bg_\mn$.
Eq.~(\ref{F17}) is a consequence of
\be
\label{F18}
W_k\left[\cJ + \cL_v \cJ\right] = W_k\left[\cJ\right]\,,
\quad \quad \cJ \equiv \left\{
  t^\mn , \sigma^\mu , \bar{\sigma}_\mu ;\, \bg_\mn \right\}\,.
\ee
This invariance property follows from (\ref{F3}) if one performs
a  compensating transformation (\ref{F9}), (\ref{F12}) on the integration 
variables $h_\mn$, $C^\mu$ and $\bar{C}_\mu$ and uses the invariance 
of $S[\bg + h],\,S_{\rm gf},\,S_{\rm gh}$ and $\Delta_k S$. 
At this point we assume that the functional measure in (\ref{F3}) 
is diffeomorphism invariant.  
     
Since the $\cR_k$'s vanish for $k=0$, the limit $k \ra 0$ of 
$\Gamma_k[g_\mn, \bg_\mn, \xi^{\mu}, \bar{\xi}_{\mu}]$ brings us 
back to the standard effective action functional which still 
depends on two metrics, though. The ``ordinary'' effective action 
$\Gamma[g_\mn]$ with one metric argument is obtained from this 
functional by setting $\bg_\mn = g_\mn$, or equivalently 
$\bar{h}_\mn =0$ \cite{back,joos}:
\be 
\Gamma[g] \equiv 
\lim_{k \ra 0} \Gamma_k[g,\bg = g, \xi =0, \bar{\xi} =0] = 
\lim_{k \ra 0} \Gamma_k[\bar{h} =0,\xi =0, \bar{\xi} =0; g = \bg]\,. 
\label{F19}
\ee   
This equation brings about the ``magic property'' of the background
field formalism: a priori the 1PI $n$-point 
functions of the metric are obtained by an $n$-fold functional 
differentiation of $\Gamma_0[\bar{h},0,0;\bg_\mn]$ with respect 
to $\bar{h}_\mn$. Hereby $\bg_\mn$ is kept fixed; it acts simply as 
an externally prescribed function which specifies the form of the 
gauge fixing condition. Hence the functional $\Gamma_0$ 
and the resulting {\it off-shell} Green functions do depend on 
$\bg_\mn$, but the {\it on-shell} Green functions, related 
to observable scattering amplitudes, do not depend on 
$\bg_\mn$. In this respect $\bg_\mn$ plays a role similar to the gauge 
parameter $\alpha$ in the standard approach. Remarkably, the same on-shell 
Green functions can be obtained by differentiating the functional 
$\Gamma[g_\mn]$ of (\ref{F19}) with respect to $g_\mn$, or 
equivalently $\Gamma_0[\bar{h} =0, \xi=0, \bar{\xi} =0;
\bg = g]$, with respect to its $\bg$ argument. In this context,
`on-shell' means that the metric satisfies the effective 
field equation $\delta \Gamma_0[g]/\delta g_\mn =0$.

With (\ref{F19}) and its $k$-dependent counterpart 
\be
\bar{\Gamma}_k[g_\mn] \equiv \Gamma_k[g_\mn, g_\mn, 0,0]\,
\label{F20}
\ee
we succeeded in constructing a diffeomorphism invariant 
generating functional for gravity: thanks to (\ref{F17}) 
$\Gamma[g_\mn]$ and $\bar{\Gamma}_k[g_\mn]$ are invariant
 under general coordinate transformations $\delta g_\mn
= \cL_v g_\mn$. However, there is  a price to be paid for their
invariance: the simplified functional $\bar{\Gamma}_k[g_\mn]$ does not 
satisfy an exact RG equation, basically because it contains insufficient
information. The actual RG evolution has to be performed 
at the level of the functional $\Gamma_k[g,\bg,\xi,\bar{\xi}\,]$.
Only {\it after} the evolution one may set $\bg = g,\, \xi =0, 
\bar{\xi} =0$. As a result, the actual theory space 
of QEG, $\{A[g,\bg,\xi,\bar{\xi} \,] \}$, consists of 
functionals of all four variables, $g_\mn, \bg_\mn, \xi^{\mu}, 
\bar{\xi}_\mu$, subject to the invariance condition (\ref{F17}).

The derivation of the FRGE for $\Gamma_k$ is analogous to the scalar 
case. Following exactly the same steps one arrives at 
\be
\label{F21}
\begin{split}
\partial_t \Gamma_k[\bar h, \xi, \bar\xi; \bg] =&
\frac{1}{2}\Tr
\left[\left(\Gamma^{(2)}_k + \widehat \cR_k\right)^{-1}_{\bar h \bar h}
\left(\partial_t \widehat \cR_k\right)_{\bar h \bar h} \right]
\\[2mm]
&-  \frac{1}{2}\Tr
\left[\left\{
 \left(\Gamma^{(2)}_k + \widehat \cR_k\right)^{-1}_{\bar \xi \xi}
-\left(\Gamma^{(2)}_k + \widehat \cR_k\right)^{-1}_{\xi \bar\xi}
\right\}
\left(\partial_t \widehat \cR_k\right)_{\bar \xi \xi} \right]\,. 
\end{split}
\ee 
Here $\Gamma^{(2)}_k$ denotes the Hessian of $\Gamma_k$ with 
respect to the dynamical fields $\bar{h},\, \xi,\,\bar{\xi}$ at fixed 
$\bg$. It is a block matrix labeled by the fields $\varphi_i \equiv
\{\bar{h}_\mn,\,\xi^\mu, \bar{\xi}_\mu\}$:
\be
\label{F22}
\Gamma^{(2)\, ij}_k(x,y) \equiv 
\frac{1}{\sqrt{\bg(x)\bg(y)}} \,
\frac{\delta^2 \Gamma_k}{\delta \varphi_i(x)\delta \varphi_j(y)}\,.
\ee
(In the ghost sector the derivatives are understood as left derivatives.)
Likewise, $\widehat{\cR}_k$ is a block diagonal matrix  with entries 
$(\widehat{\cR}_k)_{\bar{h}\bar{h}}^{\mu\nu\rho\sigma} \equiv 
\kappa^2 (\cR_k^{\rm grav}[\bg])^{\mu\nu\rho\sigma}$ and 
$\widehat{\cR}_{\bar{\xi} \xi} = \sqrt{2} \cR_k^{\rm gh}[\bg]$.
Performing the trace in the position representation it includes 
an integration $\int\! d^dx \sqrt{\bg(x)}$ involving the
background volume element. For any cutoff which is qualitatively 
similar to the exponential cutoff the traces on the RHS of eq.~(\ref{F21}) 
are well convergent, both in the IR and the UV. By virtue of the 
factor $\partial_t \widehat \cR_k$, the dominant contributions come 
from a narrow band of generalized momenta centered around $k$. Large 
momenta are exponentially suppressed.

Besides the FRGE the effective average action also satisfies  an 
exact integro-differential equation similar to (\ref{E12}) in 
the scalar case. By the same argument as there it can be used to find the $k
\ra \infty$ limit of the average action:
\be
\Gamma_{k \ra \infty}[\bar{h},\xi,\bar{\xi};\bg] = 
S[\bg+\bar{h}] + S_{\rm gf}[\bar{h};\bg] 
+S_{\rm gh}[\bar{h},\xi,\bar{\xi};\bg]\,.
\label{F24}
\ee
Note that the `initial value' $\Gamma_{k \ra \infty}$ includes the 
gauge fixing and ghost actions. At the level of the functional 
$\bar{\Gamma}_k[g]$, eq.~(\ref{F24}) boils down to 
$\bar{\Gamma}_{k \ra \infty}[g] = S[g]$. However, as $\Gamma_k^{(2)}$ 
involves derivatives with respect to $\bar{h}_\mn$ (or equivalently 
$g_\mn$) at fixed $\bg_\mn$ it is clear that the evolution cannot 
be formulated entirely in terms of $\bar{\Gamma}_k$ alone. 

The background gauge invariance of $\Gamma_k$, expressed in eq.~(\ref{F17}),
is of enormous practical importance. It implies that if the 
initial functional does not contain non-invariant terms, 
the flow will not generate such terms. Very often this reduces the 
number of terms to be retained in a reliable truncation ansatz quite 
considerably. Nevertheless, even if the initial action is simple, the RG 
flow will generate all sorts of local and non-local terms in 
$\Gamma_k$ which are consistent with the symmetries. 

Let us close this section by remarking that, at least formally,
 the construction of the effective average action 
can be repeated for Lorentzian signature metrics. In this case one deals with oscillating 
exponentials $e^{iS}$, and for arguments like the one leading to 
(\ref{F24}) one has to employ the Riemann-Lebesgue lemma. Apart from 
the obvious substitutions $\Gamma_k \ra - i \Gamma_k,\, \cR_k 
\ra - i \cR_k$ the evolution equation remains unaltered.

\section{Truncated flow equations}
\setcounter{equation}{0}
Solving the FRGE (\ref{F21}) subject to the initial condition (\ref{F24}) 
is equivalent to (and in practice as difficult as) calculating the 
original functional integral over $\gamma_\mn$. It is therefore important 
to devise efficient approximation methods. The truncation of theory 
space is the one which makes maximum use of the FRGE reformulation of the 
quantum field theory problem at hand. 

As for the flow on the theory space $\{A[g,\bg,\xi,\bar{\xi}]\}$ 
a still very general truncation consists of neglecting the evolution of the 
ghost action by making the ansatz
\be
\label{G1}
\Gamma_k[g,\bg ,\xi ,\bar\xi]
= \bar\Gamma_k[g]+\widehat\Gamma_k[g,\bg]
+S_{\rm gf}[g-\bg ;\bg] +S_{\rm gh}[g-\bg ,\xi,\bar\xi ;\bg]\,,
\ee
where we extracted the classical $S_{\rm gf}$ and $S_{\rm gh}$
from $\Gamma_k$. The remaining functional depends on both 
$g_\mn$ and $\bg_\mn$. It is further decomposed as $\bar\Gamma_k+
\widehat\Gamma_k$ where $\bar\Gamma_k$ is defined as in (\ref{F20}) 
and $\widehat\Gamma_k$ contains the deviations for $\bg\neq g$. 
Hence, by definition, $\widehat\Gamma_k[g,g]=0$, and 
$\widehat\Gamma_k$ contains in particular quantum corrections to 
the gauge fixing term which vanishes for $\bg = g$, too. 
This ansatz satisfies the initial condition (\ref{F24}) 
if
\be
\bar\Gamma_{k\ra \infty} = S \qquad \mbox{and}\qquad
\widehat \Gamma_{k\ra \infty} =0\,.
\ee
Inserting (\ref{G1}) into the exact FRGE \eqref{F21}
one obtains an evolution equation on the truncated  space 
$\{ A[g,\bg]\}$:
\ba
\label{G2}
\partial_t\Gamma_k[g,\bg]
&=& \frac{1}{2}\Tr
\left[\left(
  \kappa^{-2}\Gamma^{(2)}_k[g,\bg] +\cR_k^{\rm grav}[\bg]
      \right)^{-1}
  \partial_t \cR^{\rm grav}_k[\bg]
\right]
\nonum 
&& -\Tr\left[\left(
  -\cM[g,\bg]+ \cR^{\rm gh}_k[\bg] \right)^{-1} \dd_t 
\cR^{\rm gh}_k[\bg] \right]\,.
\ea
This equation evolves the functional 
\be
\Gamma_k[g,\bg] \equiv  \bar\Gamma_k[g]+S_{\rm gf}[g-\bg;\bg]
    +\widehat\Gamma_k[g,\bg]\,.
\ee
Here $\Gamma^{(2)}_k$ denotes the Hessian of $\Gamma_k[g, \bg]$ with respect
to $g_\mn$ at fixed $\bg_\mn$. 

The truncation ansatz (\ref{G1}) is still too general for practical 
calculations to be easily possible. The first truncation for 
which the RG flow has been found \cite{mr} is the ``Einstein-Hilbert
truncation'' which retains in $\bar{\Gamma}_k[g]$ only the terms 
$\int\! d^dx \, \sqrt{g}$ and $\int\! d^dx \, \sqrt{g} R$, already present in the  
in the classical action, with $k$-dependent coupling constants, 
and includes only the wave function renormalization in $\widehat{\Gamma}_k$:
\be
\label{G3}
\Gamma_k[g,\bg]  =  
2\kappa^2 Z_{Nk} \int\! d^dx \,\sqrt{g} \left\{ 
-R(g)+ 2\bar\lambda_k \right\}
 + \frac{ Z_{Nk}}{2 \alpha}  \int \!d^dx \,\sqrt{\bg} \, 
\bar g^\mn F_\mu F_\nu\,.
\ee
In this case the truncation subspace is 2-dimensional. The ansatz 
(\ref{G3}) contains two free functions of the scale, the running 
cosmological constant $\bar{\lb}_k$ and $Z_{Nk}$ or, equivalently,
the running Newton constant $G_k \equiv \bar{G}/Z_{Nk}$. Here $\bar{G}$ 
is a fixed constant, and $\kappa \equiv (32 \pi \bar{G})^{-1/2}$. 
As for the gauge fixing term, $F_{\mu}$ is given by eq.~(\ref{F10})     
with $\bar{h}_\mn \equiv g_\mn - \bg_\mn$ replacing $h_\mn$; it vanishes 
for $g = \bg$. The ansatz (\ref{G3}) has the general structure of 
(\ref{G1}) with $\widehat{\Gamma}_k = (Z_{Nk} -1) S_{\rm gf}$. Within the 
Einstein-Hilbert approximation the gauge fixing parameter $\alpha$ is 
kept constant. Here we shall set $\alpha =1$ and comment on generalizations 
later on.

Upon inserting the ansatz (\ref{G3}) into the flow 
equation (\ref{G2}) it boils down to a system of two ordinary differential 
equations for $Z_{Nk} $ and $\bar{\lb}_k$. Their derivation is rather 
technical, so we shall focus on the conceptual aspects here. 
In order to find $\dd_t Z_{Nk}$ and $\dd_t \bar{\lb}_k$ it is 
sufficient to consider (\ref{G2}) for $g_\mn = \bg_\mn$. In this case 
the LHS of the flow equation becomes 
$2 \kappa^2 \int \! d^dx \sqrt{g} [- R(g) \dd_t Z_{Nk} + 
2 \dd_t( Z_{Nk} \bar{\lb}_k)]$. The RHS is assumed to admit an expansion 
in terms of invariants $P_{i}[g_\mn]$. In the Einstein-Hilbert truncation only two of them, 
$\int\! d^dx \, \sqrt{g}$ and $\int\! d^dx \, \sqrt{g} R$, need to be retained.
 They can be extracted from the traces in (\ref{G2}) 
by standard derivative expansion techniques. Equating the result to 
the LHS and comparing the coefficients of $\int\! d^dx \sqrt{g}$ and $\int\! d^dx \sqrt{g}
R$, a pair of coupled differential equations for $Z_{Nk}$ and 
$\bar{\lb}_k$ arises. It is important to note that, on the RHS, we may set 
$g_\mn = \bg_\mn$ only {\it after} the functional derivatives of 
$\Gamma^{(2)}_k$ have been obtained since they must be taken at 
fixed $\bg_\mn$. 

In principle this calculation can be performed without ever considering 
any specific metric $g_\mn = \bg_\mn$. This reflects the fact that 
the approach is background covariant. The RG flow is universal 
in the sense that it does not depend on any specific metric. 
In this respect gravity is not different from the more 
traditional applications of the renormalization group: the RG 
flow in the Ising universality class, say, has nothing to do 
with any specific spin configuration, it rather reflects the 
statistical properties of very many such configurations. 

While there is no conceptual necessity to fix the background 
metric, it nevertheless is sometimes advantageous from a computational 
point of view to pick a specific class of backgrounds. Leaving        
$\bg_\mn$ completely general, the calculation of the functional traces is 
very hard work usually. In principle there exist well known 
derivative expansion and heat kernel techniques which could 
be used for this purpose, but their application is an 
extremely lengthy and tedious task usually. Moreover, typically 
the operators $\Gamma_k^{(2)}$ and $\cR_k$ are of a complicated non-standard 
type so that no efficient use of the tabulated Seeley coefficients 
can be made. However, often calculations of this type simplify if 
one can assume that $g_\mn= \bg_\mn$ has specific properties. 
Since the beta functions are background independent we may therefore 
restrict $\bg_\mn$ to lie in a conveniently chosen class of 
geometries which is still general enough to disentangle the invariants
retained and at the same time simplifies the calculation. 

For the Einstein-Hilbert truncation the most efficient choice is a 
family of $d$-spheres $S^d(r)$, labeled by their radius $r$. 
For those geometries, $D_{\alpha} R_{\mn \rho \sigma} =0$, so 
they give a vanishing value to all invariants constructed from 
$g = \bg$ containing covariant derivatives acting on curvature 
tensors. What remains (among the local invariants) are terms of the 
form $\int \! \sqrt{g} P(R)$, where $P$ is a polynomial in the 
Riemann tensor with arbitrary index contractions. To linear 
order in the (contractions of the) Riemann tensor the two 
invariants relevant for the Einstein-Hilbert truncation 
are discriminated by the $S^d$ metrics as the latter scale    
differently with the radius of the sphere: $\int \! \sqrt{g} 
\sim r^d$, $\int \! \sqrt{g} R \sim r^{d-2}$. Thus, in order 
to compute the beta functions of $\bar{\lb}_k$ and $Z_{Nk}$ 
it is sufficient to insert an $S^d$ metric with arbitrary $r$ 
and to compare  the coefficients of $r^d$ and $r^{d-2}$.      
If one wants to do better and include the three quadratic 
invariants $\int \!R_{\mn \rho\sigma} R^{\mn \rho \sigma}$, 
$\int \! R_\mn R^\mn$, and $\int \! R^2$, the family $S^d(r)$ 
is not general enough to separate them; all scale like $r^{d-4}$   
with the radius. 

Under the trace we need the operator $\Gamma_k^{(2)}[\bar{h};\bg]$.
It is most easily calculated by Taylor expanding the truncation ansatz,
$\Gamma_k[\bg + \bar{h}, \bg] = \Gamma_k[\bg,\bg] 
+ O(\bar{h}) + \Gamma_k^{\rm quad}[\bar{h};\bg] + O(\bar{h}^3)$,
and stripping off the two $\bar{h}$'s from the quadratic term, 
$\Gamma_k^{\rm quad} = \frac{1}{2} \int \! \bar{h} \Gamma_k^{(2)} \bar{h}$. 
For $\bg_\mn$ the metric on $S^d(r)$ one obtains 
\ba
\label{G4}
\Gamma_k^{\rm quad}[\bar{h};\bg]
&=& 
\frac{1}{2}Z_{Nk} \kappa^2 \int \!d^dx\, \Bigg\{
  \widehat{h}_\mn
  \left[-\bar D^2-2\bar\lambda_k+C_T\bar{R} \right]\widehat{h}^\mn
\nonum 
&&
\bspace \sspace -\left(\frac{d-2}{2d}\right)\phi
  \left[-\bar D^2-2\bar\lambda_k+C_S\bar{R} \right]\phi \Bigg\}\,,
\ea
with
$C_T \equiv (d(d-3)+4)/(d(d-1))$, $C_S \equiv (d-4)/d$.
In order to partially diagonalize this quadratic form $\bar{h}_\mn$
has been decomposed into a traceless part $\widehat{h}_\mn$ and 
the trace part proportional to $\phi$: $\bar{h}_\mn = 
\widehat{h}_\mn + d^{-1} \bg_\mn \phi$, $\bg^\mn \widehat{h}_\mn =0$. 
Further, $\bar{D}^2 = \bg^\mn \bar{D}_\mu \bar{D}_\nu$ is the 
covariant Laplace operator corresponding to the background geometry,
and $\bar{R} = d(d-1)/r^2$ is the numerical value of the curvature 
scalar on $S^d(r)$. 

At this point we can fix the constants $\cZ_k$ which appear in the 
cutoff operators $\cR_k^{\rm grav}$ and $\cR_k^{\rm gh}$ of 
(\ref{F12}). They  should be adjusted in such a way that for every 
low--momentum mode the cutoff combines with the kinetic term of 
this mode to $-\bar D^2+k^2$ times a constant. Looking at (\ref{G4}) 
we see that the respective kinetic terms for $\widehat{h}_\mn$ 
and $\phi$ differ by a factor of $-(d-2)/2d$. This suggests the following 
choice:
\be
\label{G6}
\left(\cZ_k^{\rm grav}\right)^{\mn\rho\sigma}
= \left[   \left(\one -P_\phi\right)^{\mn\rho\sigma}
 -\frac{d-2}{2d} P_\phi^{\mn\rho\sigma} \right] Z_{Nk}\,. 
\ee
Here $(P_\phi)_\mn{}^{\rho\sigma} =d^{-1} \bg_\mn\bg^{\rho\sigma}$
is the projector on the trace part of the metric.
For the traceless tensor (\ref{G6}) gives $\cZ_k^{\rm grav}=Z_{Nk} \one$, 
and for $\phi$ the different relative normalization is taken into account. (See ref.\ \cite{mr} for a detailed discussion of the subtleties related to this choice.)
Thus we obtain in the $\widehat{h}$ and the $\phi$-sector, respectively:  
\ba
\label{G7}
\left(   \kappa^{-2}\Gamma_k^{(2)}[g,g]+\cR_k^{\rm grav} 
\right)_{\widehat{h}\widehat{h}} 
\!\! &=& \!\!
Z_{Nk} \left[-D^2+k^2 R^{(0)}(-D^2/k^2)-2\bar\lambda_k+C_T R\right],
\\[2mm] 
\left( \kappa^{-2}\Gamma_k^{(2)}[g,g]+\cR_k^{\rm grav}
\right)_{\phi\phi}
\!\! &=& \!\! -\frac{d-2}{2d}
Z_{Nk} \left[-D^2+k^2 R^{(0)}(-D^2/k^2)-2\bar\lambda_k+C_S R\right]
\nonumber
\ea
From now on we may set $\bg=g$ and for simplicity we have omitted 
the bars from the metric and the curvature. Since we did not take
into account any renormalization effects in the ghost action
we set $Z_k^{\rm gh}\equiv1$ in $\cR_k^{\rm gh}$ and obtain
\be
\label{G8}
-\cM + \cR_k^{\rm gh} = - D^2 +k^2 R^{(0)}(-D^2/k^2)+C_V R\,,
\ee
with $C_V \equiv -1/d$. At this point the operator under the first trace on the RHS of 
(\ref{G2}) has become block diagonal, with the $\widehat{h} 
\widehat{h}$ and $\phi \phi$ blocks given by (\ref{G7}).
 Both block operators are expressible in terms of the 
Laplacian $D^2$, in the former case acting on traceless symmetric   
tensor fields, in the latter on scalars. The second trace in (\ref{G2}) 
stems from the ghosts; it contains (\ref{G8}) with $D^2$ acting 
on vector fields. 

It is now a matter of straightforward algebra to compute the first 
two terms in the derivative expansion of those traces, proportional 
to $\int\! d^d x\sqrt{g} \sim r^d$ and $\int\! d^d x\sqrt{g}R 
\sim r^{d-2}$. Considering the trace of an arbitrary function of the 
Laplacian, $W(-D^2)$, the expansion up to second order derivatives 
of the metric is given by 
\ba
\label{G9}
\Tr[W(-D^2)] &=& (4\pi)^{-d/2} {\rm tr}(I)
\Bigg\{ Q_{d/2}[W] \int \!d^dx \, \sqrt{g}
\nonum 
&&
\qquad\qquad\qquad
+\frac{1}{6} Q_{d/2-1}[W] \int \!d^dx\, \sqrt{g}R +O(R^2) \Bigg\}\,.
\ea
The $Q_n$'s are defined as 
\be
\label{G10}
Q_n[W] = \frac{1}{\Gamma(n)} \int_0^{\infty} dz\,z^{n-1} W(z)\,,
\ee
for $n >0$, and $Q_0[W] = W(0)$ for $n =0$. The trace ${\rm tr}(I)$ 
counts the number of independent field components. It equals 
$1,\,d,$ and $(d-1)(d+2)/2$, for scalars, vectors, and symmetric traceless 
tensors, respectively. The expansion (\ref{G9}) is easily derived
using standard heat kernel and Mellin transform techniques \cite{mr}. 

Using (\ref{G9}) it is easy to calculate the traces in (\ref{G2}) 
and to obtain the RG equations in the form $\dd_t Z_{Nk} = \cdots$
and $\dd_t (Z_{Nk} \bar{\lb}_k) = \cdots$. We shall not display them 
here since it is more convenient to rewrite them in terms of the 
dimensionless running cosmological constant and Newton constant, respectively:
\be \label{G11}
\lb_k \equiv k^{-2} \bar{\lb}_k\,, \qquad
 g_k \equiv k^{d-2} G_k \equiv k^{d-2} Z_{Nk}^{-1} \bar{G}\,.
\ee  
Recall that the dimensionful running Newton constant 
is given by $G_k = Z_{Nk}^{-1} \bar{G}$.    
In terms of the dimensionless couplings $g$ and $\lb$ the RG equations
become a system of autonomous differential equations:
\begin{subeqnarray}
\dd_t g_k \, \is \, \big[d - 2 + \eta_N(g_k, \lambda_k) \big]\, g_k \equiv \beta_g(g_k, \lb_k) \, , 
\\
\dd_t \lb_k \, \is \, \beta_{\lb}(g_k, \lb_k) \,.
\label{G13}
\end{subeqnarray}
Here $\eta_N \equiv - \dd_t \ln Z_{Nk}$ is the anomalous dimension of the 
operator $\sqrt{g} R$, 
\be 
\eta_N(g_k, \lambda_k) = \frac{g_k \, B_1(\lb_k)}{ 1- g_k \, B_2(\lb_k)} \,,
\label{G14}
\ee
with the following functions of $\lb_k$:   
\ba
\label{G15}
B_1(\lambda_k)
& \equiv & \frac{1}{3}(4\pi)^{1-d/2} \Bigg[  d(d+1) \Phi^1_{d/2-1}(-2\lambda_k)
-6d(d-1)\Phi^2_{d/2}(-2\lambda_k)
\nonum 
&& \qquad\qquad \quad
-4d\Phi^1_{d/2-1}(0)-24\Phi^2_{d/2}(0)\Bigg]
\\
B_2(\lambda_k) &\equiv& -\frac{1}{6}(4\pi)^{1-d/2}
\left[  d(d+1) \widetilde\Phi^1_{d/2-1}(-2\lambda_k)
-6d(d-1)\widetilde\Phi^2_{d/2}(-2\lambda_k) \right]. 
\nonumber
\ea
The beta function for $\lb$ is given by a similar expression:
\ba
\label{G16}
\beta_{\lb}(g_k, \lambda_k) \! \! &=& \! -(2-\eta_N)\lambda_k+
\frac{1}{2} g_k(4\pi)^{1-d/2}\cdot \\
&&  \quad \cdot\left[
2d(d+1) \Phi^1_{d/2}(-2\lambda_k)-8d \Phi^1_{d/2}(0)
-d(d+1)\eta_N\widetilde\Phi^1_{d/2}(-2\lambda_k)\right]. \; \; \; \; \; \; \; \;
\nonumber  
\ea
The ``threshold functions'' $\Phi$ and $\widetilde{\Phi}$ appearing  in (\ref{G15}) 
and (\ref{G16}) are certain integrals involving the normalized 
cutoff function $R^{(0)}$: 
\ba
\label{G17}
\Phi^p_n(w) &\equiv& \frac{1}{\Gamma(n)}\int_0^\infty dz\,z^{n-1}
\frac{R^{(0)}(z)-z R^{(0)\,\prime}(z)}{[z+R^{(0)}(z)+w]^p}\,,
\nonum
\widetilde\Phi^p_n(w) &\equiv& \frac{1}{\Gamma(n)}\int_0^\infty dz\,
z^{n-1} \frac{R^{(0)}(z)}{[z+R^{(0)}(z)+w]^p}\,.
\ea 
They are defined for positive integers $p$, and $n >0$. 

With the derivation of the system (\ref{G13}) we managed to find an
approximation to a two-dimensional projection of the RG flow. 
Its properties, and in particular the domain of applicability 
and reliability of the Einstein-Hilbert truncation  will be discussed 
in the following section.

While there are (few) aspects of the truncated RG flow which are 
independent of the cutoff scheme, i.e., independent of the 
function $R^{(0)}$, the explicit solution of the flow equation 
requires a specific choice of this function. As we discussed 
already, the normalized cutoff function $R^{(0)}(w),\, 
w = p^2/k^2$, describes the ``shape'' of $\cR_k(p^2)$ 
in the transition region where it interpolates between the 
prescribed behavior for $p^2 \ll k^2$ and $p^2 \gg k^2$, 
respectively, and is referred to as the ``shape function'' therefore.
In the literature various forms of $R^{(0)}$'s have been 
employed. Easy to handle, but disadvantageous for high 
precision calculations is the sharp cutoff \cite{frank1} 
defined by $\cR_k(p^2) = \lim_{\hat{R} \ra \infty} \hat{R} 
\,\theta(1 - p^2/k^2)$, where the limit is to be taken after 
the $p^2$ integration. This cutoff allows for an evaluation 
of the $\Phi$ and $\widetilde{\Phi}$ integrals in closed form. 
Taking $d=4$ as an example, eqs.~(\ref{G13}) boil down to the 
following simple system of equations:%
\footnote{To be precise, (\ref{G18}) corresponds to the sharp 
cutoff with $s=1$, see \cite{frank1}.}
\begin{subeqnarray} 
\dd_t \lb_k \is -(2 - \eta_N) \lb_k - \frac{g_k}{\pi} 
\Big[ 5\ln(1 - 2 \lb_k) - 2 \zeta(3) + \frac{5}{2} \eta_N\Big]\,,
\\
\dd_t  g_k \is (2 + \eta_N) \,  g_k\,,
\\
\eta_N \is - \frac{2 \, g_k}{ 6\pi + 5 \, g_k} 
\Big[ \frac{18}{1 - 2 \lb_k} +  5\ln(1 - 2 \lb_k) - 
\zeta(2) + 6 \Big]\,.
\label{G18}
\end{subeqnarray}
Also the ``optimized cutoff'' \cite{opt} with 
$R^{(0)}(w) = (1-w) \theta(1-w)$ allows for an analytic 
evaluation of the integrals \cite{litimgrav}. In order to 
check the scheme (in)dependence of the results it is 
desirable to perform the calculation for a 
whole class of $R^{(0)}$'s. For this purpose the following one 
parameter family of exponential cutoffs has been used 
\cite{souma,oliver1,oliver2}: 
\be 
R^{(0)}(w;s) = \frac{sw}{e^{sw} -1}\,.
\label{G19}
\ee 
The precise form of the cutoff is controlled by the ``shape 
parameter'' $s$. For $s=1$, (\ref{G19}) coincides with the standard 
exponential cutoff. The exponential cutoffs are 
suitable for precision calculations, but the price to be  
paid is that their $\Phi$ and $\widetilde{\Phi}$ integrals 
can be evaluated only numerically. The same is true for 
a one-parameter family of shape functions with compact support 
which was used in \cite{oliver1,oliver2}. 

Above we illustrated the general ideas and constructions underlying 
gravitational RG flows by means of the simplest example, the 
Einstein-Hilbert truncation. In the literature various extensions 
have been investigated. The derivation and analysis of these more 
general flow equations, corresponding to higher dimensional 
truncation subspaces, is an extremely complex and calculationally 
demanding problem in general. For this reason we cannot go into 
the technical details here and just mention some further developments. 

\noindent
{\bf (1)} The natural next step beyond the Einstein-Hilbert 
truncation consists in generalizing the functional 
$\bar{\Gamma}_k[g]$, while keeping the gauge fixing and 
ghost sector classical, as in (\ref{G1}). During the 
RG evolution the flow generates all possible diffeomorphism invariant terms 
in $\bar{\Gamma}_k[g]$ which one can construct from 
$g_\mn$. Both local and non-local 
terms are induced. The local invariants contain strings of 
curvature tensors and covariant derivatives acting upon them, with 
any number of tensors and derivatives, and of all possible index 
structures. The first truncation of this class which has been 
worked out completely \cite{oliver2,oliver3} is the 
``$R^2$-truncation'' defined by (\ref{G1}) with the same 
$\widehat{\Gamma}_k$ as before, and the $({\rm curvature})^2$ 
action 
\be 
\bar{\Gamma}_k[g] = \int\! d^dx \sqrt{g} 
\Big\{ (16 \pi G_k)^{-1} [-R(g) + 2 \bar{\lb}_k] 
+ \bar{\beta}_k R^2(g) \Big\} \, .  
\label{G20}
\ee  
In this case the truncated theory space is 3-dimensional.
Its natural (dimensionless) coordinates are $(g, \lb ,\beta)$, 
where $\beta_k \equiv k^{4-d} \bar{\beta}_k$, and $g$ and 
$\lb$ defined in \eqref{G11}. Even though (\ref{G20}) 
contains only one additional invariant, the derivation 
of the corresponding RG equations is far more complicated 
than in the Einstein-Hilbert case. We shall summarize the 
results obtained with (\ref{G20}) \cite{oliver2,oliver3} 
in the next section.

\noindent
{\bf (2)} As for generalizing the ghost sector of the truncation 
beyond (\ref{G1}) no results are available yet, but there is 
a partial result concerning the gauge fixing term. 
Even if one makes the ansatz (\ref{G3}) for $\Gamma_k[g,\bg]$ 
in which the gauge fixing term has the classical (or more 
appropriately, bare) structure one should treat its prefactor 
as a running coupling: $\alpha = \alpha_k$. The beta function of $\alpha$ 
has not been determined yet from the FRGE, but there is 
a simple argument which allows us to bypass this calculation. 

In non-perturbative Yang-Mills theory and in perturbative 
quantum gravity $\alpha = \alpha_k =0$ is known to be a fixed point 
for the $\alpha$ evolution. The following reasoning suggests 
that the same is true within the non-perturbative FRGE approach 
to gravity. In the standard functional integral the limit 
$\alpha \ra 0$ corresponds to a sharp implementation of the 
gauge fixing condition, i.e., $\exp(-S_{\rm gf})$ becomes 
proportional to $\delta[F_{\mu}]$. The domain of the $\int \! 
\cD h_\mn $ integration consists of those $h_\mn$'s 
which satisfy the gauge fixing condition exactly, $F_\mu =0$.
Adding the IR cutoff at $k$ amounts to suppressing some of the 
$h_\mn$ modes while retaining the others. But since all of them 
satisfy $F_\mu =0$, a variation of $k$ cannot change the domain 
of the $h_\mn$ integration. The delta functional $\delta[F_\mu]$ 
continues to be present for any value of $k$ if it was there
originally. As a consequence, $\alpha$ vanishes for all $k$, 
i.e., $\alpha =0$ is a fixed point of the $\alpha$ evolution \cite{alphafp}. 

Thus we can mimic the dynamical treatment of a running $\alpha$ 
by setting the gauge fixing parameter to the constant value
$\alpha =0$. The calculation for $\alpha =0$ is more complicated 
than at $\alpha =1$, but for the Einstein-Hilbert truncation 
the $\alpha$-dependence of $\beta_g$ and $\beta_\lb$, for 
arbitrary constant $\alpha$ has been found in \cite{falkenberg,
oliver1}. The $R^2$-truncations could be analyzed only in the 
simple $\alpha=1$ gauge, but the results from the Einstein-Hilbert 
truncation suggest the UV quantities of interest do not change 
much between $\alpha =0$ and $\alpha =1$ \cite{oliver1,oliver2}. 

\noindent
{\bf (3)} Up to now we considered pure gravity. As for as the general 
formalism, the inclusion of matter fields is straightforward. 
The structure of the flow equation remains unaltered, except that 
now $\Gamma_k^{(2)}$ and $\cR_k$ are operators on the larger Hilbert
space of both gravity and matter fluctuations. In practice the 
derivation of the projected RG equations can be quite a 
formidable task, however, the difficult part being the 
decoupling of the various modes (diagonalization of 
$\Gamma_k^{(2)}$) which in most calculational schemes is 
necessary for the computation of the functional traces. 
Various matter systems, both interacting and non-interacting 
(apart from their interaction with gravity) have been studied 
in the literature \cite{percadou,grandamat,pires}. A rather 
detailed analysis has been performed 
by Percacci et al. In \cite{percadou,perper1} arbitrary multiplets of free 
(massless) fields with spin $0,1/2,1$ and $3/2$ were included. 
In  \cite{perper1} an interacting scalar theory coupled 
to gravity in the Einstein-Hilbert approximation was analyzed,
and a possible solution to the triviality and the hierarchy problem \cite{hier}
was proposed in this context.

\noindent
{\bf (4)} Finally we mention another generalization of the simplest 
case reviewed above which is of a more technical nature \cite{oliver1}. In order 
to facilitate    the calculation of the functional traces it is 
helpful to employ a transverse-traceless (TT) decomposition of the 
metric: $h_\mn = h_\mn^T + \bar{D}_\mu V_\nu + \bar{D}_\nu V_\mu 
+ \bar{D}_\mu \bar{D}_\nu \sigma - 
d^{-1} \bg_\mn \bar{D}^2 \sigma + d^{-1} \bg_\mn \phi$.    
Here $h^T_\mn$ is a transverse traceless tensor, $V_\mu$ a 
transverse vector, and $\sigma$ and $\phi$ are scalars. 
In this framework it is natural to formulate the cutoff in terms of 
the component fields appearing in the TT decomposition: 
$\Delta_k S \sim \int \! h^T_\mn \cR_k {h^T}^\mn + 
\int \! V_\mu \cR_k V^\mu + \cdots$. This cutoff is referred to 
as a cutoff of ``type B'', in contradistinction to the ``type A''
cutoff described above, $\Delta_k S \sim \int \! h_\mn \cR_k h^\mn$. 
Since covariant derivatives do not commute the two cutoffs 
are not exactly equal even if they contain the same shape function. 
Thus, comparing type A and type B cutoffs is an additional possibility 
for checking scheme (in)dependence \cite{oliver1,oliver2}. 

\section{Asymptotic Safety}
\setcounter{equation}{0}
In intuitive terms, the basic idea of asymptotic safety can be understood as follows.
The boundary of theory space depicted in fig.\ \ref{theoryspace} is meant to separate points with coordinates $\{u_\alpha, \alpha = 1,2,\cdots\}$ with all the essential couplings $u_\alpha$ well defined, from points with undefined, divergent couplings. The basic task of renormalization theory consists in constructing an ``infinitely long'' RG trajectory which lies entirely within this theory space, i.e., a trajectory which neither leaves theory space (that is, develops divergences) in the UV limit $k \rightarrow \infty$ nor in the IR limit $k \rightarrow 0$. Every such trajectory defines one possible quantum theory.

The idea of asymptotic safety is to perform the UV limit $k \rightarrow \infty$ at a fixed point $\{u_\alpha^*, \alpha = 1,2,\cdots\} \equiv u^*$ of the RG flow. The fixed point is a zero of the vector field $\vec \beta \equiv (\beta_\alpha)$, i.e., $\beta_\alpha(u^*) = 0$ for all $\alpha = 1,2,\cdots$. The RG trajectories, solutions of $k \partial_k u_\alpha(k) = \beta_\alpha(u(k))$, have a low ``velocity'' near a fixed point because the $\beta_\alpha$'s are small there and directly at the fixed point the running stops completely. As a result, one can ``use up'' an infinite amount of  RG time near/at the fixed point if one bases the quantum theory on a trajectory which runs into such a fixed point for $k \rightarrow \infty$. This is the key idea of asymptotic safety: If in the UV limit the trajectory ends at a fixed point, an ``inner point'' of theory space giving rise to a well behaved action functional, we can be sure that, for $k \rightarrow \infty$, the trajectory does not escape from theory space, i.e., does not develop pathological properties such as divergent couplings. For $k \rightarrow \infty$ the resulting quantum theory is ``asymptotically safe''  from unphysical divergences. In the context of gravity, Weinberg \cite{wein} proposed to use a non-Gaussian fixed point (NGFP) for letting $k \rightarrow \infty$. By definition, not all of its coordinates $u^*_\alpha$ 
vanish.\footnote{In contrast, $u^*_\alpha = 0, \forall \alpha = 1,2,\cdots$ is a so-called Gaussian fixed point (GFP). In a sense standard perturbation theory takes the $k \rightarrow \infty$ limit at the GFP; see \cite{livrev} for a detailed discussion.}

Recall from section 2.2 that the coordinates $u_\alpha$ are the {\it dimensionless} essential couplings related to the dimensionful ones $\bar{u}_\alpha$ by $ u_\alpha \equiv k^{-d_{\alpha}} \bar{u}_\alpha$. Hence the running of the $\bar{u}$'s is given by
\be
\ub_\alpha(k) = k^{d_{\alpha}} \, u_\alpha(k) \,.
\ee
Therefore, even directly at a NGFP where $u_\alpha(k) \equiv u^*_\alpha$, the dimensionful couplings keep running according to a power law involving their canonical dimensions $d_\alpha$:
\be
\ub_\alpha(k) = u_\alpha^* \, k^{d_{\alpha}} \,.
\ee
Furthermore, non-essential dimensionless couplings are not required to attain fixed point values.
\renewcommand{\baselinestretch}{1}
\small\normalsize
\begin{figure}[t]
\leavevmode
\begin{center}
\epsfxsize=10cm
\epsfysize=4.5cm
\epsfbox{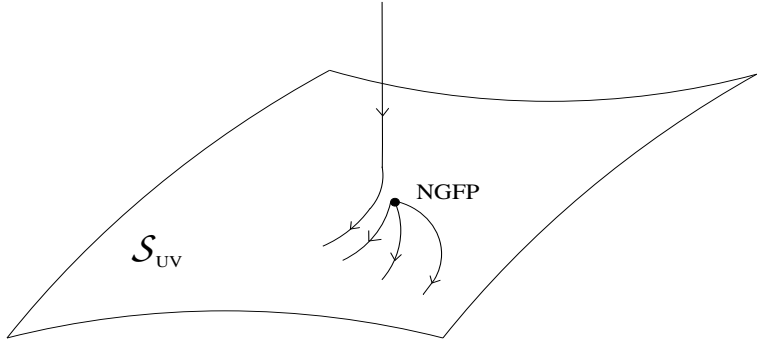}
\end{center}
\caption{\small Schematic picture of the UV critical hypersurface $\cS_{\rm UV}$ of the NGFP. It is spanned by RG trajectories emanating from the NGFP as the RG scale $k$ is lowered. Trajectories not in the surface are attracted towards $\cS_{\rm UV}$ as $k$ decreases. (The arrows point in the direction of decreasing $k$, from the ``UV'' to the ``IR''.)}
\label{UVsurface}
\end{figure}
\renewcommand{\baselinestretch}{1.5}
\small\normalsize

Given a NGFP, an important concept is its {\it UV critical hypersurface} $\cS_{\rm UV}$, or synonymously, its {\it unstable manifold}. By definition, it consists of all points of theory space which are pulled into the NGFP by the inverse RG flow, i.e., for {\it in}creasing $k$.
Its dimensionality ${\rm dim}\left({\cal S}_{\rm UV}\right)\equiv \Delta_{\rm UV}$
is given by the number of attractive (for {\it in}creasing cutoff $k$) 
directions in the space of couplings.

Writing the RG equations as
$k\,\partial_k {u}_\alpha = \beta_\alpha(u_1,u_2,\cdots)$,
the linearized flow near the fixed point is governed by the Jacobi matrix
${\bf B}=(B_{\alpha \gamma})$, $B_{\alpha \gamma}\equiv\partial_\gamma  \beta_\alpha(u^*)$:
\begin{eqnarray}
\label{H2}
k\,\partial_k\,{u}_\alpha(k)=\sum\limits_\gamma B_{\alpha \gamma}\,\left(u_\gamma(k)
-u_{\gamma}^*\right)\;.
\end{eqnarray}
The general solution to this equation reads
\begin{eqnarray}
\label{H3}
u_\alpha(k)=u_{\alpha}^*+\sum\limits_I C_I\,V^I_\alpha\,
\left(\frac{k_0}{k}\right)^{\theta_I}
\end{eqnarray}
where the $V^I$'s are the right-eigenvectors of ${\bf B}$ with eigenvalues 
$-\theta_I$, i.e., $\sum_{\gamma} B_{\alpha \gamma}\,V^I_\gamma =-\theta_I\,V^I_\alpha$. Since ${\bf B}$ is not symmetric in general the $\theta_I$'s are not guaranteed to be real. We
assume that the eigenvectors form a complete system though. Furthermore, $k_0$ 
is a fixed reference scale, and the $C_I$'s are constants of integration. 

If $u_\alpha(k)$ is to describe a trajectory in $\cS_{\rm UV}$,
 $u_\alpha(k)$ must 
approach $u_{\alpha}^*$ in the limit
$k\rightarrow\infty$ and therefore we must set $C_I=0$ for all $I$ with 
${\rm Re}\,\theta_I<0$. Hence the dimensionality $\Delta_{\rm UV}$ equals the 
number of ${\bf B}$-eigenvalues with a negative real part, i.e., the number of
$\theta_I$'s with ${\rm Re}\,\theta_I >0$. The corresponding eigenvectors 
span the tangent space to
$\cS_{\rm UV}$ at the NGFP.

If $u_\alpha(k)$ describes a generic trajectory with all
$C_I$ nonzero and we {\it lower} the cutoff, only $\Delta_{\rm UV}$ 
``relevant'' parameters
corresponding to the eigendirections tangent to $\cS_{\rm UV}$ grow 
(${\rm Re}\, \theta_I > 0$), while the remaining ``irrelevant'' couplings 
pertaining to the eigendirections normal to $\cS_{\rm UV}$ decrease 
(${\rm Re}\, \theta_I < 0$). Thus near the NGFP a generic trajectory 
is attracted towards $\cS_{\rm UV}$, see fig.\ \ref{UVsurface}.

Coming back to the asymptotic safety construction, let us now 
use this fixed point in order to take the limit $k \ra \infty$. 
The trajectories which define an infinite cutoff limit for QEG are 
special in that all irrelevant couplings are set to zero: $C_I = 0$ 
if ${\rm Re} \, \theta_I < 0$. These conditions place the trajectory 
exactly on $\cS_{\rm UV}$. There is a $\Delta_{\rm UV}$-parameter family 
of such trajectories, and the experiment must decide which one is 
realized in Nature. 
Therefore the predictive power of the theory increases with decreasing
 dimensionality of ${\cal S}_{\rm UV}$, i.e., number of UV attractive eigendirections of the
NGFP. (If $\Delta_{\rm UV} < \infty$, the quantum field 
theory thus constructed is comparable to and as predictive as a perturbatively
renormalizable model with $\Delta_{\rm UV}$ ``renormalizable couplings'', i.e.,
couplings relevant at the GFP.)


The quantities $\theta_I$ are referred to as critical exponents 
since when the renormalization group is applied to critical 
phenomena (second order phase transitions) the traditionally 
defined critical exponents are related to the $\theta_I$'s in a 
simple way \cite{avactrev}. In fact, one of the early successes 
of the RG ideas was an explanation of the universality properties
of critical phenomena, i.e., the fact that systems 
at the critical point seem to ``forget'' the precise form 
of their microdynamics and just depend on the universality 
class, characterized by a set of critical exponents, they 
belong to. 

In the present context, ``universality'' means that
 certain, very special, quantities related to the RG flow are
 independent of the precise form of
the cutoff and, in particular, its shape function $R^{(0)}$. 
Universal quantities are potentially measurable or at least 
closely related to observables. The $\theta_I$'s are examples of universal 
quantities, while the coordinates of the fixed point, 
$u_{\alpha}^*$, are not, even in an exact calculation. 
Quantities independently known to be universal provide an 
important tool for testing the reliability or accuracy of 
{\it approximate} RG calculations and of truncations in particular.  
Since they are known to be $R^{(0)}$ independent in an 
exact treatment, we can determine the degree of their 
$R^{(0)}$-dependence within the truncation and use it 
as a measure for the quality of the truncated calculation. 

For a more detailed and formal discussion of asymptotic safety and, in particular, its relation to perturbation theory we refer to the review \cite{livrev}.
\section{Average Action approach to Asymptotic Safety}
\setcounter{equation}{0}
Our discussion of the asymptotic safety construction in the 
previous section was at the level of the exact (untruncated)
RG flow. In this section we are going to implement these ideas in the context
of explicitly computable approximate RG flows on truncated theory spaces. 
We shall mostly concentrate on the Einstein-Hilbert (``$R$--'') and the $R^2$--truncation of pure gravity in $d=4$. The corresponding $d$-dimensional flow equations were derived in refs.\ \cite{mr} and \cite{oliver2}, respectively.

\subsection{The phase portrait of the Einstein-Hilbert truncation}

\begin{figure}[t]
\leavevmode
\hskip 14mm
\epsfxsize=13cm
\epsfysize=8.9cm
\epsfbox{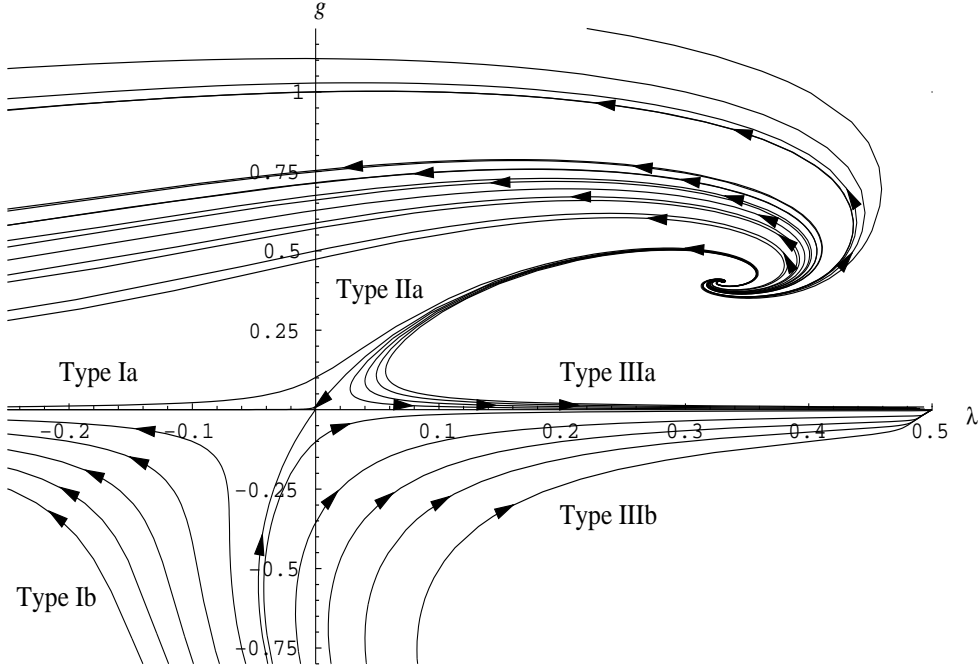}
\vskip 2mm
\caption{\small
RG flow in the $g$-$\lb-$plane. The arrows point in the direction 
of increasing coarse graining, i.e., of decreasing $k$. (From \cite{frank1}.)}
\label{fig0}
\end{figure}

In \cite{frank1} the RG equations (\ref{G13}) implied by the 
Einstein-Hilbert truncation have been analyzed in detail, using both 
analytical and numerical methods. In particular all RG trajectories
of this system have been classified, and examples
have been computed numerically. The most important classes of 
trajectories in the phase portrait on the $g$-$\lb-$plane 
are shown in fig.~\ref{fig0}. The trajectories were obtained 
by numerically solving the system (\ref{G18}) for a 
sharp cutoff; using a smooth one all qualitative features 
remain unchanged.    
The RG flow is found to be dominated by two fixed points $(g^*, \lb^*)$:
the GFP at $g^* = \lb^* =0$, and 
a NGFP with $g^* > 0$ and 
$\lb^* >0$. There are three classes of trajectories emanating from the NGFP:
trajectories of Type Ia and IIIa run towards negative and positive
cosmological constants, respectively, and the single trajectory of
Type IIa (``separatrix'') hits the GFP for $k\to 0$. The
high momentum properties of QEG are governed by the 
NGFP; for $k \to \infty$, in fig.\ \ref{fig0} 
all RG trajectories on the half--plane
$g>0$ run into this point. Note that near the NGFP the dimensionful Newton constant vanishes 
for $k \ra \infty$ according to $G_k \equiv g_k/k^2 \approx g^*/k^2
\ra 0$, while the cosmological constant diverges: $\bar{\lambda}_k \equiv \lambda_k k^2 \approx \lambda^* k^2 \rightarrow \infty$. 
%
\renewcommand{\baselinestretch}{1}
\small\normalsize
\renewcommand{\baselinestretch}{1.5}
\small\normalsize
%

So, the Einstein-Hilbert truncation does indeed predict the existence of a NGFP
with exactly the properties needed for the asymptotic safety construction. Clearly the crucial question to be analyzed now is whether the NGFP found is the projection of a fixed point in the exact theory on the untruncated theory space or whether it is merely the artifact of an insufficient approximation.

\subsection{Testing the Einstein-Hilbert truncation}
We mentioned already that the residual $R^{(0)}$-dependence of universal quantities is a 
measure for the quality of a truncation. This test has been applied to 
the Einstein-Hilbert truncation
in \cite{souma,oliver1}. 
 We shall display the results in the next subsection.
In accordance with the general       
theory the coordinates of the fixed point $(g^*, \lb^*)$ 
are not universal. However, it can be argued that they should 
give rise to a universal combination, the product $g^* \lb^*$ which
 can be measured in principle \cite{oliver1}. 
While $k$ and, at a fixed value of $k$, $G_k$ and $\bar{\lb}_k$
cannot be measured separately, we may invert the function $k \mapsto 
G_k$ and insert the result $k = k(G)$ into $\bar{\lb}_k$. This 
leads to an in principle experimentally testable relationship 
$\bar{\lb} = \bar{\lb}(G)$ between Newton's constant and the 
cosmological constant. Here $\bar{\lb}$ and $G$ should be determined
in experiments involving similar scales. In the fixed point regime 
this relationship reads $\bar{\lb}(G) = g^* \lb^*/G$. So, 
even if this is quite difficult in practice, one can determine
the product $g^* \lb^*$ experimentally. As a consequence 
in any reliable calculation $g^* \lb^*$ should be approximately 
$R^{(0)}$ independent.

The ultimate justification of a given truncation consists in 
checking that if one adds further terms to it, its physical 
predictions remain robust. The first step towards testing the 
robustness of the Einstein-Hilbert truncation near the NGFP
against the inclusion of other invariants has been taken in 
refs.~\cite{oliver2,oliver3} where the $R^2$--truncation of 
eq.~(\ref{G20}) has been analyzed. The corresponding beta 
functions for the three generalized couplings $g, \lb$ 
and $\beta$ have been derived, but they are too complicated to be 
reproduced here. Suffice it to say that on the 
3-dimensional $(g,\lb,\beta)$ space, too, a NGFP has been
found which generalizes the one from the pure $R$--calculation.     
This allows for a comparison of the fixed point results for 
the $R^2$-- and the Einstein-Hilbert truncation, and for a check
of the approximate $R^{(0)}$ independence of universal 
quantities in the 3-dimensional setting. 
For the Einstein-Hilbert truncation the universality analysis  
has been performed for an arbitrary constant gauge parameter 
$\alpha$, including the `physical' value $\alpha =0$ \cite{oliver1}.  Because of its 
algebraic complexity the $R^2$--analysis \cite{oliver2} 
has been carried out in the simpler $\alpha=1$ gauge.  

\subsection{Evidence for Asymptotic Safety}

We now summarize the results concerning the NGFP which 
were obtained with the $R$-- (items (1)-(5)) and $R^2$--truncation (items (6)-(9)), respectively
\cite{oliver1,frank1,oliver2,oliver3}. All properties mentioned below
are independent pieces of evidence pointing in the direction 
that QEG is indeed asymptotically safe in 
four dimensions. Except for point (5) all results refer to $d=4$.  

\noindent
{\bf (1) Universal existence:} 
Both for type A and type B cutoffs the non-Gaussian fixed point exists for all
shape functions $R^{(0)}$. (This generalizes earlier results in
\cite{souma}.) It seems impossible to find an admissible cutoff which
destroys the fixed point in $d=4$. This result is highly non-trivial since in
higher dimensions $(d\gtrsim 5)$ the existence of the NGFP depends on the cutoff 
chosen \cite{frank1}.

\noindent
{\bf (2) Positive Newton constant:}
While the position of the fixed point is scheme dependent, all cutoffs yield
{\it positive} values of $g^*$ and $\lambda^*$. A negative $g^*$ might have 
been problematic for stability reasons, but there is no mechanism in the flow
equation which would exclude it on general grounds. 

\noindent
{\bf (3) Stability:}
For any cutoff employed the NGFP is found to be UV
attractive in both directions of the $\lambda$-$g-$plane. Linearizing the
flow equation according to eq. (\ref{H2}) we obtain a pair of complex
conjugate critical exponents $\theta_1=\theta_2^*$ with positive real part 
$\theta'$ and imaginary parts $\pm\theta''$. In terms of $t = \ln(k/k_0)$
the general solution to the linearized flow equations reads
\begin{eqnarray}
\label{H4}
\left(\lambda_k,g_k\right)^{\bf T}
&=&\left(\lambda^*,g^*\right)^{\bf T}
+2\Bigg\{\left[{\rm Re}\,C\,\cos\left(\theta''\,t\right)
+{\rm Im}\,C\,\sin\left(\theta''\,t\right)\right]
{\rm Re}\,V\nonumber\\
& &+\left[{\rm Re}\,C\,\sin\left(\theta''\,t\right)-{\rm Im}\,C
\,\cos\left(\theta''\,t\right)\right]{\rm Im}\,V\Bigg\}e^{-\theta' t}\;.
\end{eqnarray}
with $C\equiv C_1=(C_2)^*$ an arbitrary complex number and $V\equiv 
V^1=(V^{2})^*$ the right-eigenvector of ${\bf B}$ with eigenvalue $-\theta_1
=-\theta_2^*$. Eq.~(\ref{H2}) implies that, due to the positivity of 
$\theta'$, all trajectories hit the fixed point as $t$ is sent to infinity. 
The non-vanishing imaginary part $\theta''$ has no impact on the stability. 
However, it influences the shape of the trajectories which spiral into the 
fixed point for $k\rightarrow\infty$. Thus, the fixed 
point has the stability properties needed in the asymptotic safety scenario.

Solving the full, non-linear flow equations \cite{frank1} shows that the 
asymptotic scaling region where the linearization (\ref{H4}) is valid 
extends from $k=\infty$ down to about $k\approx m_{\rm Pl}$ with the
Planck mass defined as $m_{\rm Pl}\equiv G_0^{-1/2}$. Here $m_{\rm Pl}$ 
plays a role similar to $\Lambda_{\rm QCD}$ in QCD: it marks the lower 
boundary of the asymptotic scaling region. We set $k_0\equiv
m_{\rm Pl}$ so that the asymptotic scaling regime extends from about 
$t=0$ to $t=\infty$.

\noindent
{\bf (4) Scheme- and gauge dependence:}
Analyzing the cutoff scheme dependence of $\theta'$, $\theta''$, and 
$g^*\lambda^*$ as a measure for the reliability of the truncation,
the critical exponents were found to be reasonably constant within about a 
factor of 2. For $\alpha=1$ and
$\alpha=0$, for instance, they assume values in the ranges $1.4\lesssim
\theta'\lesssim 1.8$, $2.3\lesssim\theta''\lesssim 4$ and $1.7\lesssim
\theta'\lesssim 2.1$, $2.5\lesssim\theta''\lesssim 5$, respectively. The
universality properties of the product $g^*\lambda^*$ are even more
impressive. Despite the rather strong scheme dependence of $g^*$ and 
$\lambda^*$ separately, their product has almost no visible $s$-dependence for
not too small values of $s$. Its value is
\begin{eqnarray}
\label{H5}
g^*\lambda^*\approx\left\{\begin{array}{l}\mbox{$0.12$ for $\alpha=1$}\\
\mbox{$0.14$ for $\alpha=0$\,.}\end{array}\right.
\end{eqnarray}
The difference between the ``physical'' (fixed point) value of the gauge
parameter, $\alpha=0$, and the technically more convenient $\alpha=1$ are at 
the level of about 10 to 20 percent. 

\noindent
{\bf (5) Higher and lower dimensions:} The beta functions implied 
by the FRGE are continuous functions of the spacetime dimensionality and it is 
instructive to analyze them for $d \neq 4$. 
In ref.\ \cite{mr} it has been shown that for $d = 2 + \epsilon$, $|\epsilon| \ll 1 $, the FRGE reproduces Weinberg's \cite{wein} fixed point for Newton's constant, $g^* = \frac{3}{38}\epsilon$, and also supplies a corresponding fixed point value for the cosmological constant, $\lambda^* = - \frac{3}{38} \Phi^1_1(0) \epsilon$, with the treshold function given in \eqref{G17}. 
For arbitrary $d$ and a generic cutoff the RG flow is quantitatively 
similar to the 4-dimensional one for all $d$ smaller than a certain 
critical dimension $d_{\rm crit}$, above which the existence or non-existence of the NGFP becomes  cutoff-dependent. The critical dimension is scheme dependent, but for any admissible 
cutoff it lies well above $d=4$. As $d$ approaches $d_{\rm crit}$ from 
below, the scheme dependence of the universal quantities increases 
drastically, indicating that the $R$-truncation becomes insufficient 
near $d_{\rm crit}$. 

 In fig.~\ref{fixd} we show 
the $d$-dependence of $g^*$, $\lb^*$, $\th'$, and $\th''$ for 
two versions of the sharp cutoff (with $s=1$ and $s=30$, respectively)
and for the exponential cutoff with $s=1$. For $2 + \eps \leq d \leq 4$
the scheme dependence of the critical exponents is rather weak;
it becomes appreciable only near $d \approx 6$ \cite{frank1}. 
Fig.~\ref{fixd} suggests that the Einstein-Hilbert 
truncation in $d=4$ performs almost as well as near $d=2$. Its validity can be extended towards larger dimensionalities by optimizing the shape function \cite{litimgrav}. 

\renewcommand{\baselinestretch}{1}
\small\normalsize
\begin{figure}[t]
\renewcommand{\baselinestretch}{1}
\epsfxsize=0.49\textwidth
\begin{center}
\leavevmode
\epsffile{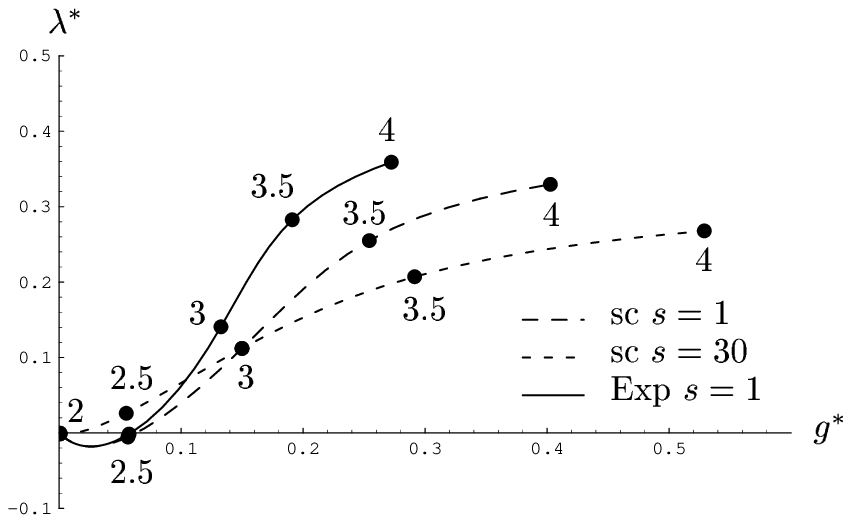}
\epsfxsize=0.48\textwidth
\epsffile{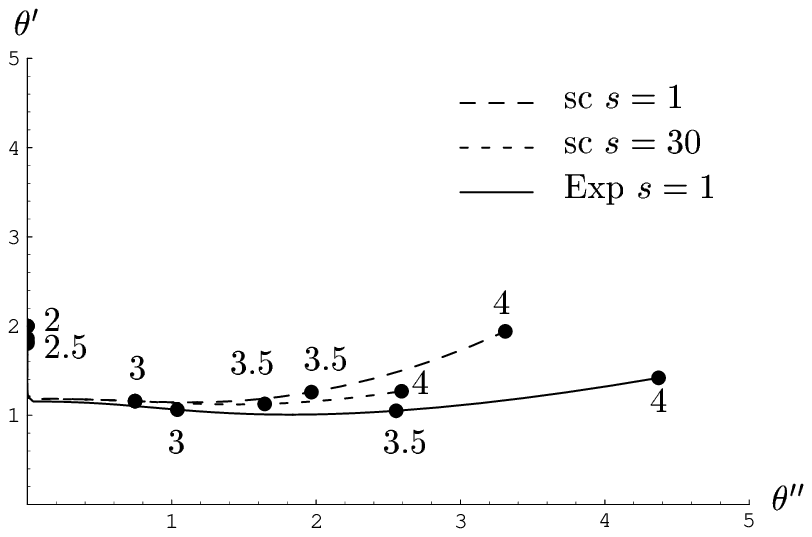}
\end{center}
\begin{center}
\leavevmode
\epsfxsize=0.6\textwidth
\epsffile{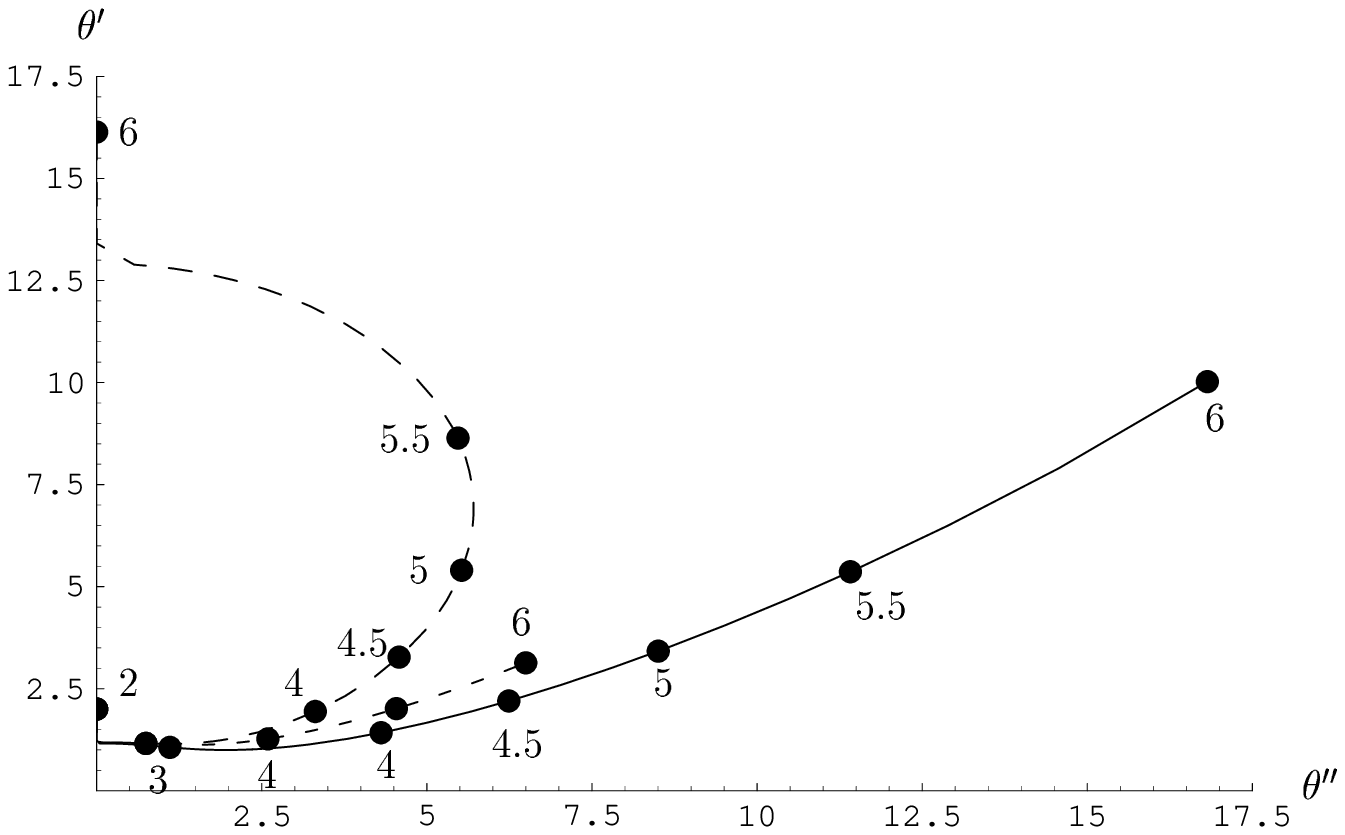}
\end{center}
\parbox[c]{\textwidth}{\caption{\label{fixd}{\footnotesize Comparison of $\lambda^*,g^*, 
\theta^{\prime}$ and $\theta^{\prime \prime }$ for different cutoff 
functions in dependence of the dimension $d$. Two versions of the sharp 
cutoff (sc) and the exponential cutoff with $s=1$ (Exp) have been employed.
 The upper line shows that for $2+ \epsilon \le d \le 4$ the cutoff scheme dependence of the
 results is rather small. The lower diagram shows that increasing $d$ beyond 
about 5 leads to a significant difference in the results for $\theta^{\prime}, 
\theta^{\prime \prime}$ obtained with the different cutoff schemes. (From \cite{frank1}.)}}}
\end{figure}
\renewcommand{\baselinestretch}{1.5}
\small\normalsize

\begin{figure}[t]
\begin{minipage}{7.9cm}
        \epsfxsize=7.9cm
        \epsfysize=5.2cm
        \centerline{\epsffile{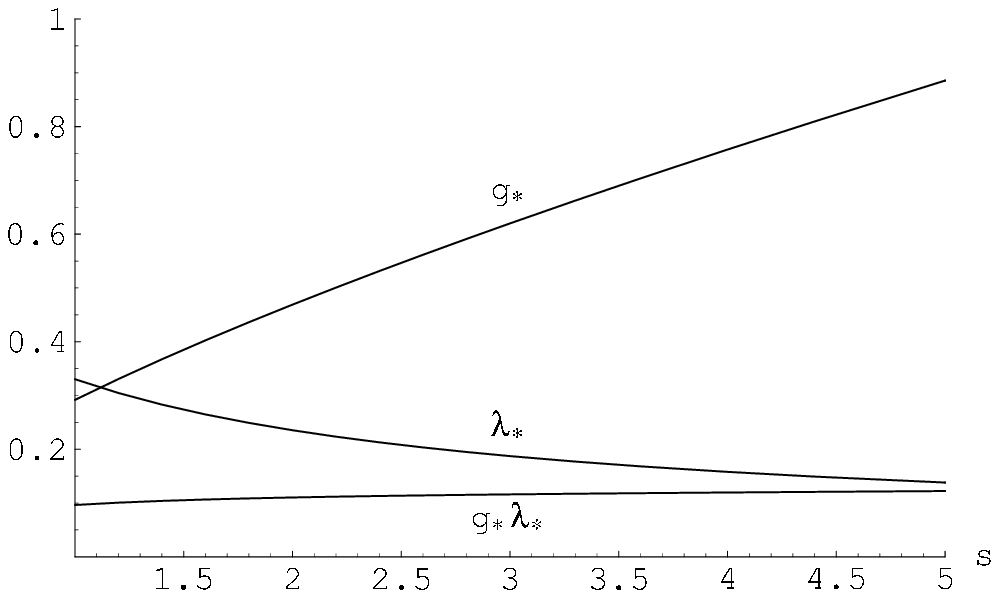}}
\centerline{(a)}
\end{minipage}
\hfill
\begin{minipage}{7.5cm}
        \epsfxsize=7.5cm
        \epsfysize=5cm
        \centerline{\epsffile{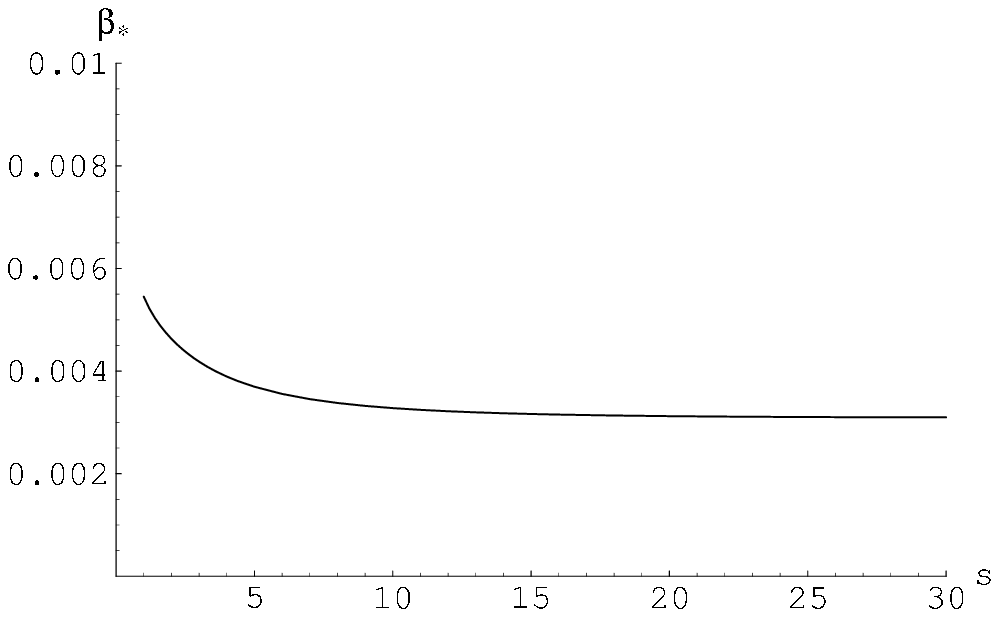}}
\centerline{(b)}
\end{minipage}
\vspace{0.3cm}
\caption{(a) $g^*$, $\lambda^*$, and $g^*\lambda^*$ as functions of $s$ 
for $1\le s\le 5$, and (b) $\beta^*$ as a function of $s$ for $1\le s\le 30$,
using the family of exponential shape functions (\ref{G19}). 
(From ref.~\cite{oliver3}.)}  
\label{plot1}
\end{figure}

\noindent
{\bf (6)  Position of the fixed point $(R^2)$:}
Also with the generalized truncation the NGFP is found to exist for all
admissible cutoffs. Fig.\ \ref{plot1} shows its coordinates 
$(\lambda^*,g^*,\beta^*)$ for the family of shape functions (\ref{G19}) and
the type B cutoff. For every shape parameter $s$, the values of $\lambda^*$ 
and $g^*$ are almost the same as those obtained with the Einstein-Hilbert 
truncation. In particular, the product $g^*\lambda^*$ is constant with a very
high accuracy. For $s=1$, for instance, one obtains 
$(\lambda^*,g^*)=(0.348,0.272)$ from the Einstein-Hilbert truncation and
$(\lambda^*,g^*,\beta^*)=(0.330,0.292,0.005)$ from the generalized truncation.
It is quite remarkable that $\beta^*$ is always significantly
smaller than $\lambda^*$ and $g^*$. Within the limited precision of our
calculation this means that in the 3-dimensional parameter space the fixed
point practically lies on the $\lambda$-$g-$plane with $\beta=0$, 
i.e., on the parameter space of the pure Einstein-Hilbert truncation.

\renewcommand{\baselinestretch}{1}
\small\normalsize
\renewcommand{\baselinestretch}{1.5}
\small\normalsize

\noindent
{\bf (7) Eigenvalues and -vectors $(R^2)$:}
The NGFP of the
$R^2$-truncation proves to be UV attractive in any of the three directions of
the $(\lambda,g,\beta)$ space for all cutoffs used. The linearized flow in
its vicinity is always governed by a pair of complex conjugate critical
exponents $\theta_1=\theta'+{\rm i}\theta''=\theta_2^*$ with $\theta'>0$ and
a single real, positive critical exponent $\theta_3>0$. It may be expressed as
\begin{eqnarray}
\label{H6}
\left(\lambda_k,g_k,\beta_k\right)^{\bf T}
&=&\left(\lambda^*,g^*,\beta^*\right)^{\bf T}
+2\Bigg\{\left[{\rm Re}\,C\,\cos\left(\theta''\,t\right)
+{\rm Im}\,C\,\sin\left(\theta''\,t\right)\right]
{\rm Re}\,V\nonumber\\
& &+\left[{\rm Re}\,C\,\sin\left(\theta''\,t\right)-{\rm Im}\,C
\,\cos\left(\theta''\,t\right)\right]{\rm Im}\,V\Bigg\}\,e^{-\theta' t}
+C_3 V^3\,e^{-\theta_3 t} \, \qquad
\end{eqnarray}
with arbitrary complex $C\equiv C_1=(C_2)^*$ and real $C_3$, and
with $V\equiv V^1=(V^2)^*$ and $V^3$ the right-eigenvectors of the stability
matrix $(B_{ij})_{i,j\in\{\lambda,g,\beta\}}$ with eigenvalues $-\theta_1=
-\theta_2^*$ and $-\theta_3$, respectively. Clearly the conditions for UV 
stability are $\theta'>0$ and $\theta_3>0$. They are indeed satisfied for all 
cutoffs. For the exponential shape function with $s=1$, for instance, we find
$\theta'=2.15$, $\theta''=3.79$, $\theta_3=28.8$, and 
${\rm Re}\,V=(-0.164,0.753,-0.008)^{\bf T}$,
${\rm Im}\,V=(0.64,0,-0.01)^{\bf T}$, $V^3=-(0.92,0.39,0.04)^{\bf T}$. (The
vectors are normalized such that $\left\|V\right\|=\left\|V^3\right\|=1$.)
The trajectories (\ref{H6}) comprise three independent normal modes with
amplitudes proportional to ${\rm Re}\,C$, ${\rm Im}C$ and $C_3$, respectively.
The first two are again of the spiral type while the third one is a straight line.

For any cutoff, the numerical results have several quite remarkable
properties. They all indicate that, close to the NGFP, the
RG flow is rather well approximated by the pure Einstein-Hilbert truncation.

\noindent
{\bf (a)} The $\beta$-components of ${\rm Re}\,V$ and ${\rm Im}\,V$ are very
tiny. Hence these two vectors span a plane which virtually coincides with the
$g$-$\lambda-$subspace at $\beta=0$, i.e., with the parameter space of the
Einstein-Hilbert truncation. As a consequence, the  ${\rm Re}\,C$-- and
${\rm Im}C$--normal modes are essentially the same trajectories as the
``old'' normal modes already found without the $R^2$--term. Also the
corresponding $\theta'$-- and $\theta''$--values coincide within the scheme
dependence.

\noindent
{\bf (b)}
The new eigenvalue $\theta_3$ introduced by the $R^2$--term is
significantly larger than $\theta'$. When a trajectory approaches the fixed
point from below $(t\rightarrow\infty)$, the ``old'' normal modes $\propto
{\rm Re}\,C,{\rm Im}\,C$ are proportional to $\exp(-\theta' t)$, but the new
one is proportional to $\exp(-\theta_3 t)$, so that it decays much quicker.
 For every trajectory running into the fixed point, i.e., for every
set of constants $({\rm Re}\,C,{\rm Im}\,C,C_3)$, we find therefore that once
$t$ is sufficiently large the trajectory lies entirely in the
${\rm Re}\,V$-${\rm Im}\,V-$subspace, i.e., the $\beta=0$-plane practically.

Due to the large value of $\theta_3$, the new scaling field is very
``relevant''. However, when we start at the fixed point $(t=\infty)$
and lower $t$ it is only at the low energy scale $k\approx m_{\rm Pl}$
$(t\approx 0)$ that $\exp(-\theta_3 t)$ reaches unity, and only then, i.e.,
far away from the fixed point, the new scaling field starts growing
rapidly.

\noindent
{\bf (c)}
Since the matrix ${\bf B}$ is not symmetric its eigenvectors have
no reason to be orthogonal. In fact, one finds that $V^3$ lies almost in the
${\rm Re}\,V$-${\rm Im}\,V-$plane. For the angles between the eigenvectors
given above we obtain $\sphericalangle ({\rm Re}\,V,{\rm Im}\,V)=102.3^\circ$, 
$\sphericalangle({\rm Re}\,V,V^3)=100.7^\circ$, 
$\sphericalangle({\rm Im}\,V,V^3)=156.7^\circ$. Their sum is $359.7^\circ$
which confirms that ${\rm Re}\,V$, ${\rm Im}\,V$ and $V^3$ are almost 
coplanar. This implies that when we lower $t$ and move away from the fixed
point so that the $V^3$--scaling field starts growing, it is again
predominantly the $\int d^dx\,\sqrt{g}$ and $\int\! d^dx\,\sqrt{g}R$
invariants which get excited, but not $\int\! d^dx\,\sqrt{g}R^2$ in 
the first place.

\begin{figure}[t]
\begin{minipage}{7.9cm}
        \epsfxsize=7.9cm
        \epsfysize=5.2cm
        \centerline{\epsffile{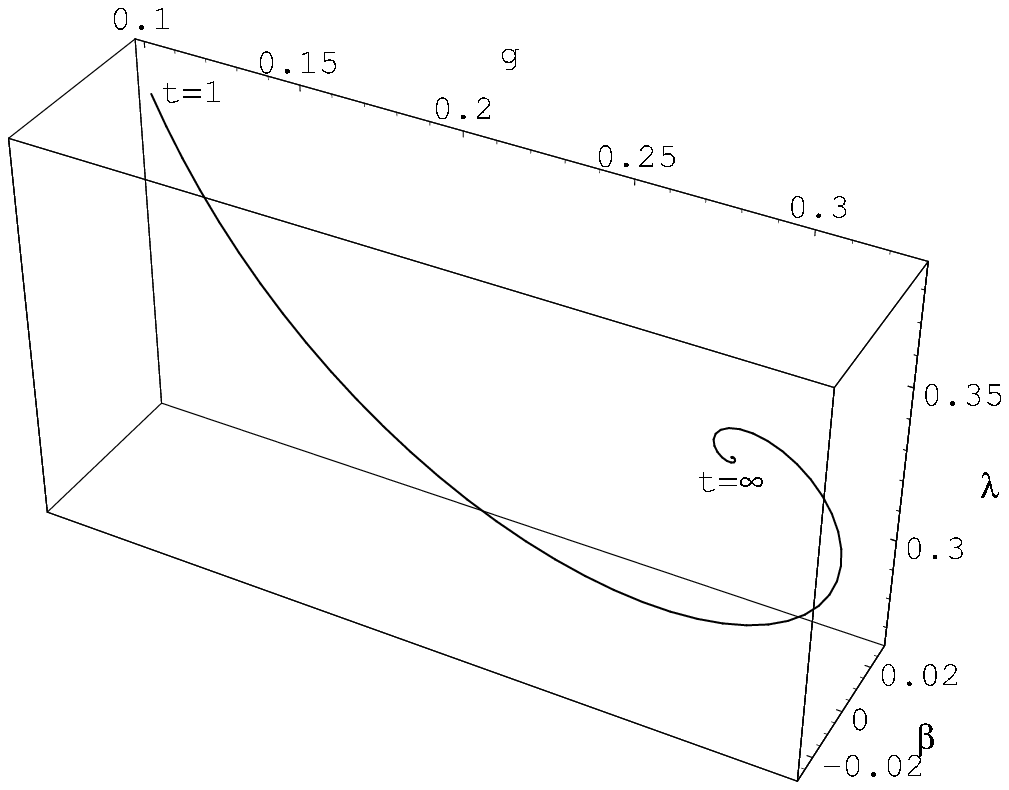}}
\centerline{(a)}
\end{minipage}
\hfill
\begin{minipage}{7.9cm}
        \epsfxsize=7.9cm
        \epsfysize=5.2cm
        \centerline{\epsffile{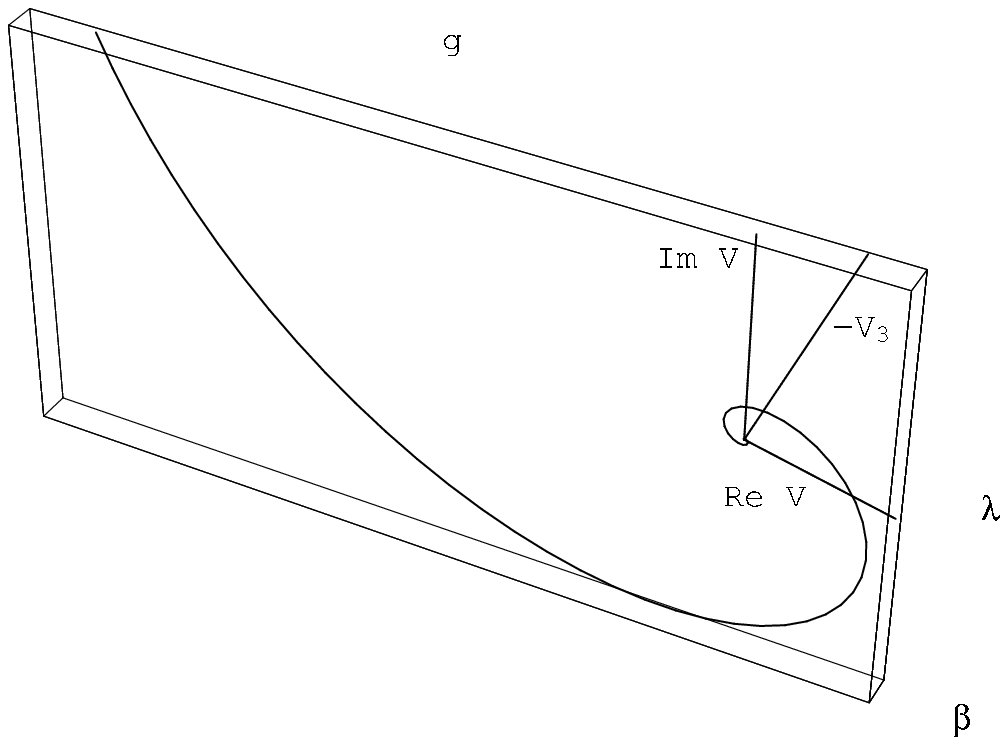}}
\centerline{(b)}
\end{minipage}
\vspace{0.3cm}
\caption{Trajectory of the linearized flow equation obtained from the
$R^2$--truncation for $1\le t=\ln(k/k_0)<\infty$. In (b) we depict
the eigendirections and the ``box'' to which the trajectory is 
confined. (From ref.~\cite{oliver3}.)}  
\label{plot2}
\end{figure}

Summarizing the three points above, we can say that very close to the fixed
point the RG flow seems to be essentially two-dimensional, and that this
two-dimensional flow is well approximated by the RG equations of the
Einstein-Hilbert truncation. In fig.\ \ref{plot2} we show a typical trajectory
which has all three normal modes excited with equal strength
$({\rm Re}\,C={\rm Im}\,C=1/\sqrt{2}$, $C_3=1)$. All the way down from
$k=\infty$ to about $k=m_{\rm Pl}$ it is confined to a very thin
box surrounding the $\beta=0$--plane.

\renewcommand{\baselinestretch}{1}
\small\normalsize
\renewcommand{\baselinestretch}{1.5}
\small\normalsize

\begin{figure}[t]
\begin{minipage}{7.9cm}
        \epsfxsize=7.9cm
        \epsfysize=5.2cm
       \centerline{\epsffile{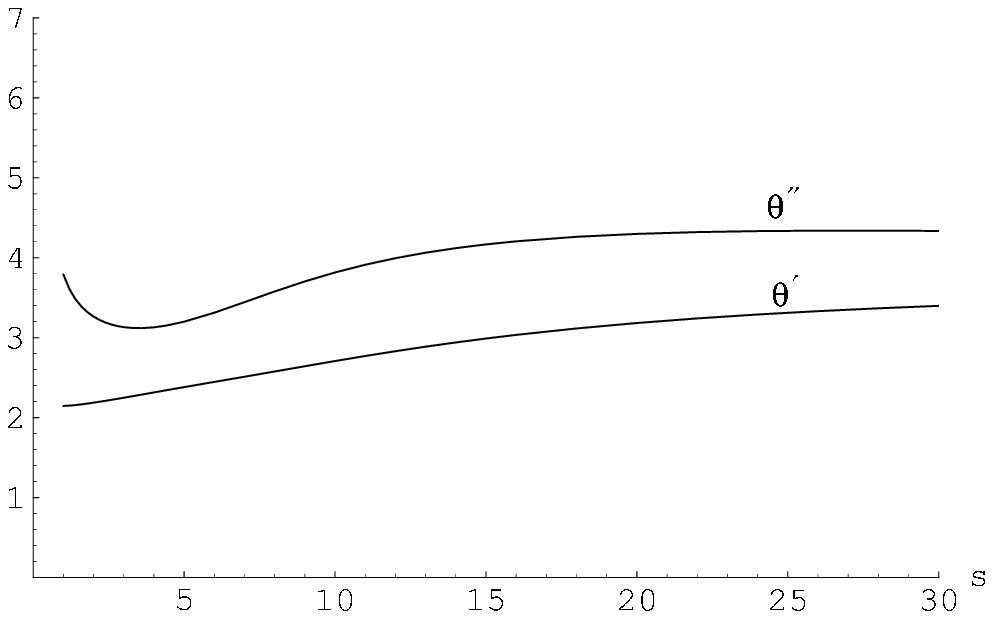}}
\centerline{(a)}
\end{minipage}
\hfill
\begin{minipage}{7.9cm}
        \epsfxsize=7.9cm
        \epsfysize=5.2cm
    \centerline{\epsffile{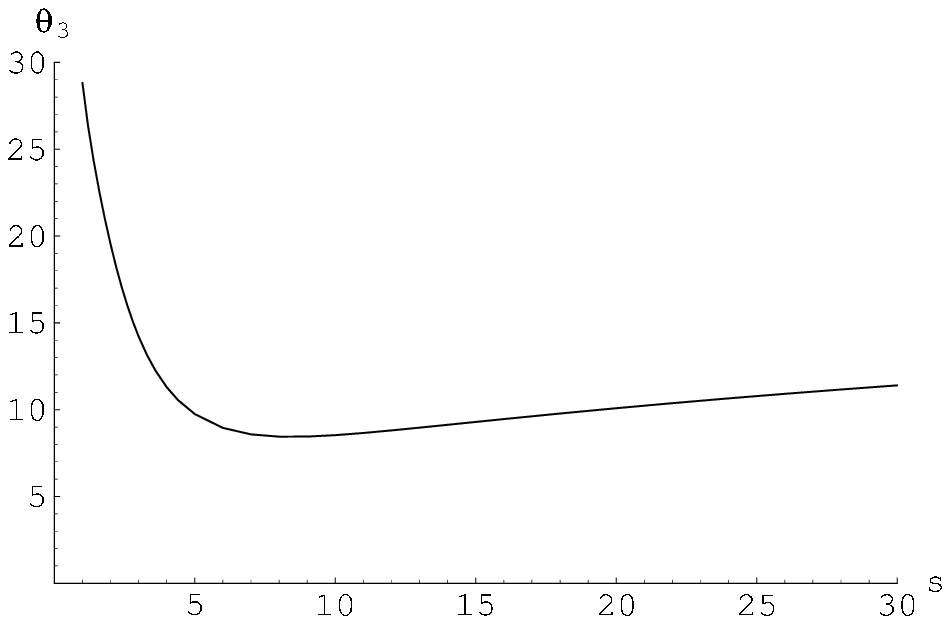}}
\centerline{(b)}
\end{minipage}
\vspace{0.3cm}
\caption{(a) $\theta'={\rm Re}\,\theta_1$ and $\theta''={\rm Im}\,\theta_1$, 
and (b) $\theta_3$ as functions of $s$, using the family of exponential shape 
functions (\ref{G19}). (From \cite{oliver2}.)}  
\label{plot3}
\end{figure}

\noindent
{\bf (8) Scheme Dependence $(R^2)$:}
The scheme dependence of the critical exponents and of the product
$g^*\lambda^*$ turns out to be of the same order of magnitude as in the case
of the Einstein-Hilbert truncation. Fig.\ \ref{plot3} shows the cutoff
dependence of the critical exponents, using the family of shape functions
(\ref{G19}). For the cutoffs employed $\theta'$ and $\theta''$ assume values
in the ranges $2.1\lesssim\theta'\lesssim 3.4$ and $3.1\lesssim\theta''
\lesssim 4.3$, respectively. While the scheme dependence of $\theta''$ is
weaker than in the case of the Einstein-Hilbert truncation one finds that it is
slightly larger for $\theta'$. The exponent $\theta_3$ suffers from relatively
strong variations as the cutoff is changed, $8.4\lesssim\theta_3\lesssim
28.8$, but it is always significantly larger than $\theta'$.
The product $g^*\lambda^*$ again exhibits an extremely weak scheme dependence.
Fig. \ref{plot1}(a) displays $g^*\lambda^*$ as a function of $s$. 
It is impressive to see how the cutoff
dependences of $g^*$ and $\lambda^*$ cancel almost perfectly. Fig.
\ref{plot1}(a) suggests the universal value $g^*\lambda^*\approx 0.14$.
Comparing this value to those obtained from the Einstein-Hilbert truncation
we find that it differs slightly from the one based upon
the same gauge $\alpha=1$. The deviation is of the same size as the difference
between the $\alpha=0$-- and the $\alpha=1$--results of the Einstein-Hilbert
truncation.
\renewcommand{\baselinestretch}{1}
\small\normalsize
\renewcommand{\baselinestretch}{1.5}
\small\normalsize

As for the universality of the critical exponents we emphasize that the
qualitative properties listed above (e.g., $\theta',\theta_3>0$, $\theta_3\gg
\theta'$, etc.) obtained for all cutoffs. The $\theta$'s have a
much stronger scheme dependence than $g^*\lambda^*$, however. This is most
probably due to neglecting further relevant operators in the truncation so
that the ${\bf B}$-matrix we are diagonalizing is too small still.

\noindent
{\bf (9) Dimensionality of ${\cal S}_{\rm UV}$:}
According to the canonical dimensional analysis, the
(curvature)$^n$-invariants in 4 dimensions are classically 
marginal for
$n=2$ and irrelevant for $n>2$. The results for $\theta_3$ indicate that there
are large non-classical contributions so that there might be relevant operators
perhaps even beyond $n=2$. With the present approach it is clearly not
possible to determine their number $\Delta_{\rm UV}$. However, as it is
hardly conceivable that the quantum effects change the signs of arbitrarily
large (negative) classical scaling dimensions, $\Delta_{\rm UV}$ should be
finite \cite{wein}. 

A first confirmation of this picture comes from the $R^2$-calculation
which has also been performed in $d=2+\varepsilon$ where, at least 
canonically,  the dimensional count is shifted by two units. In
this case we find indeed that the third scaling field is irrelevant,
$\theta_3<0$. Therefore the dimensionality of
${\cal S}_{\rm UV}$ could be as small as $\Delta_{\rm UV}=2$, but this is not
a proof, of course. If so, the quantum theory would be characterized by only
two free parameters, the renormalized Newton constant and cosmological constant, respectively.

\section{Discussion and Conclusion}
\setcounter{equation}{0}

On the basis of the above results we believe that the non-Gaussian 
fixed point occurring in the Einstein-Hilbert truncation is not a 
truncation artifact but rather the projection of a fixed
point in the exact theory space. The fixed point and all its qualitative 
properties are stable against variations of the cutoff
and the inclusion of a further invariant in the truncation. It is particularly
remarkable that within the scheme dependence the additional $R^2$--term has
essentially no impact on the fixed point. We interpret the above results and
their mutual consistency as quite non-trivial indications supporting the
conjecture that 4-dimensional QEG indeed possesses a RG
fixed point with precisely the properties needed for its non-perturbative
renormalizability and asymptotic safety. 

Recently this picture has been beautifully confirmed by Codello, Percacci and Rahmede \cite{r6}
who, in $d=4$, considered truncations of the form
\be
\bar{\Gamma}_k[g] = \int d^4x \sqrt{g} \, \sum_{n=0}^N \bar{u}_n(k) \, R^n \, .
\ee
In the most advanced case the highest power of the curvature scalar was as large as $N=7$. An important result obtained with these truncations is that going beyond the $R^2$ truncation the new eigendirections at the NGFP are all UV repulsive (Re\,$\theta_I < 0$), indicating that $\Delta_{\rm UV}$ is indeed likely to be a small finite number. Increasing the order $N$ of the curvature polynomial the values of the universal quantities show a certain degree of convergence, in particular $g^*\lambda^*$ agrees with the Einstein-Hilbert result \eqref{H5} to within 10 or 20 percent for any $N=2, \cdots, 7$. It is quite amazing how well the RG flow near the NGFP is approximated by the Einstein-Hilbert truncation; the reason for this is not yet fully understood.

In these notes we focused on the average action approach to QEG. For a detailed discussion including evidence for asymptotic safety from other approaches we refer to \cite{livrev}.

Before closing, some further comments might be helpful here. \\
%
{\bf (1)} The construction of an effective average action for gravity as introduced in \cite{mr} represents a {\it background independent} approach to quantum gravity. Somewhat paradoxically, this background independence is achieved by means of the background field formalism: One fixes an arbitrary background, quantizes the fluctuation field in this background, and afterwards adjusts $\bar{g}_{\mu \nu}$ in such a way that the expectation value of the fluctuation vanishes: $\bar{h}_{\mu \nu} = 0$. In this way the background gets fixed dynamically. 

\noindent
{\bf (2)} The combination of the effective average action with the background field method has been successfully tested within conventional field theory. In QED and Yang-Mills type gauge theories it reproduces the known results and extends them into the non-perturbative domain \cite{ym,ymrev}.

\noindent
{\bf (3)} The coexistence of asymptotic safety and perturbative non-renormalizability is well understood. In particular upon fixing $\bg_{\mu \nu} = \eta_{\mu \nu}$ and expanding the trace on its RHS in powers of $G$ the FRGE reproduces the divergences of perturbation theory; see ref.\ \cite{livrev} for a detailed discussion of this point.

\noindent
{\bf (4)} It is to be emphasized that in the average action framework the RG flow, i.e., the vector field $\vec{\beta}$, is completely determined once a theory space is fixed. As a consequence, the choice of theory space determines the set of fixed points $\Gamma^*$ at which asymptotically safe theories can be defined. Therefore, in the asymptotic safety scenario the bare action $S = \Gamma^*$ is a {\it prediction} of the theory rather than an ad hoc postulate as usually in quantum field theory. (Ambiguities could arise only if there is more than one suitable NGFP.)

\noindent
{\bf (5)} According to the results available to date, the Einstein-Hilbert action of classical General Relativity seems not to play any distinguished role in the asymptotic safety context, at least not at the conceptual level. The only known NGFP on the theory space of QEG has the structure $\Gamma^* = \mbox{Einstein-Hilbert action} + \mbox{``more''}$ where ``more'' stands for both local and non-local corrections. So it seems that the Einstein-Hilbert action is only an approximation to the true fixed point action, albeit an approximation which was found to be rather reliable for many purposes.

\noindent
{\bf (6)} Any quantum theory of gravity must reproduce the successes of classical General Relativity. As for QEG, it cannot be expected that this will happen for all RG trajectories in $\cS_{\rm UV}$, but it should happen for some or at least one of them. Within the Einstein-Hilbert truncation it has been shown \cite{h3} that there actually do exist trajectories (of type IIIa) which have an extended classical regime and are consistent with all observations.

\noindent
{\bf (7)} In the classical regime mentioned above the spacetime geometry is non-dynamical to a very good approximation. In this regime the familiar methods of quantum field theory in curved classical spacetimes apply, and it is clear therefore that effects such as Hawking radiation or cosmological particle production are reproduced by the general framework of QEG with matter.

\noindent
{\bf (8)} Coupling free massless matter fields to gravity, 
it turned out \cite{perper1} that the fixed point continues to exist under very 
weak conditions concerning the number of various types 
of matter fields (scalars, fermions, etc.). No fine tuning 
with respect to the matter multiplets is necessary. In particular 
 asymptotic safety does not seem to require 
any special constraints or symmetries among the matter fields such as 
supersymmetry, for instance. 

\noindent
{\bf (9)} Since the NGFP seems to exist already in pure gravity it is 
likely  that a widespread prejudice about 
gravity may be incorrect: its quantization seems not to  
require any kind of unification with the other fundamental 
interactions.  
\vspace{8mm}

Given the situation that by now the asymptotic safety of QEG hardly can be questioned any more, future work will have to focus on its physics implications. The effective average action is an ideal framework for investigations of this sort since, contrary to other exact RG schemes, it provides a  family of scale dependent {\it effective} (rather than {\it bare}) actions, $\{ \Gamma_k[ \, \cdot \, ], 0 \le k < \infty \}$. Dealing with phenomena involving typical scales $k$, a tree--level evaluation of $\Gamma_k$ is sufficient for finding the leading quantum gravity effects. The investigations  already performed in this direction employed the following methods.

\noindent
{\bf (a)} {\it RG improvement}: In refs.\ \cite{bh} and \cite{cosmo1}, respectively, a first study of the asymptotic safety-based ``phenomenology'' of black hole and cosmological spacetimes has been carried out by ``RG improving'' the classical field equations or their solutions. Hereby $k$ is identified with a fixed, geometrically motivated scale. Using the same method, modified dispersion relations of point particles were discussed in \cite{girelli}. 

\noindent
{\bf (b)} {\it Scale dependent geometry:} In the spirit of the gravitational average action, a spacetime manifold can be visualized as a fixed differentiable manifold equipped with infinitely many metric structures $\{ \langle g_{\mu \nu} \rangle_k , 0 \le k < \infty \}$ where $\langle g_{\mu \nu} \rangle_k$ is a solution to the effective field equation implied by $\Gamma_k$. Comparable to the situation in fractal geometry the metric, and therefore all distances, depend on the resolution of the experiment by means of which spacetime is probed. A general discussion of the geometrical issues involved (scale dependent diffeomorphisms, symmetries, causal structures, etc.) was given in \cite{jan2}, and in \cite{jan1} these ideas were applied to show that QEG can generate a minimum length dynamically. In \cite{oliver1,oliver2} it has been pointed out that the QEG spacetimes should have fractal properties, with a fractal dimension equal to 4 on macroscopic and 2 on microscopic scales. This picture was confirmed by the computation of their spectral dimension in \cite{oliverfrac}. Quite remarkably, the same dynamical dimensional reduction from 4 to 2 has also been observed in Monte-Carlo simulations using the causal triangulation approach \cite{ajl1,ajl2,ajl34}. It is therefore intriguing to speculate that this discrete approach and the gravitational average action actually describe the same underlying theory.

\noindent
{\bf Acknowledgements}\\
F.S.\ is supported by the European Commission Marie Curie Fellowship no.\
MEIF-CT-2005-023966.

\end{document}